\pdfoutput=1
\documentclass[PAPER, atlasdraft=false, cernpreprint=true, texlive=2023, UKenglish,orcidlogo, texmf, orcidlogo]{atlasdoc}
\usepackage[full]{atlaspackage}
\usepackage{atlasbiblatex}
 
\usepackage{atlasphysics}

\graphicspath{{logos/}{figures/}{aux_figures/}}
\usepackage{ANA-STDM-2018-35-PAPER-defs}

\AtlasTitle{Measurements of electroweak $W^{\pm}Z$  boson pair production in association with two jets in $pp$ collisions at $\sqrt{s} = 13$~TeV with the ATLAS detector}
 
\AtlasAbstract{
Measurements of integrated and differential cross-sections for electroweak $W^{\pm}Z$ production in association with two jets ($W^{\pm}Zjj$) in proton--proton collisions are presented.
The data collected by the ATLAS detector at the Large Hadron Collider from $2015$ to $2018$ at a centre-of-mass energy of $\sqrt{s} = 13$~TeV are used, corresponding to an integrated luminosity of $140$~fb$^{-1}$.
The $W^{\pm}Zjj$ candidate events are reconstructed using leptonic decay modes of the gauge bosons.
Events containing three identified leptons, either electrons or muons, and two jets are selected.
Processes involving pure electroweak $W^{\pm}Zjj$ production at Born level are separated from $W^{\pm}Zjj$ production involving a strong coupling.
The measured integrated fiducial cross-section of electroweak $W^{\pm}Zjj$ production per lepton flavour is $\sigma_{WZjj\mathrm{-EW} \rightarrow \ell^{'} \nu \ell \ell jj} = 0.368 \; \pm  0.037 \,(\mathrm{stat.}) \; \pm 0.059  \,(\mathrm{syst.}) \; \pm 0.003 \,(\mathrm{lumi.})  \; \mathrm{fb}$, where $\ell$ and $\ell^{'}$ are either an electron or a muon.
Respective cross-sections of electroweak and strong $W^{\pm}Zjj$ production are measured separately for events with exactly two jets or with more than two jets, and in three bins of the invariant mass of the two jets.
The inclusive $W^{\pm}Zjj$ production cross-section, without separating electroweak and strong production, is also measured to be  $\sigma_{WZjj  \rightarrow \ell^{'} \nu \ell \ell jj} = 1.462 \; \pm 0.063 \,(\mathrm{stat.}) \; \pm 0.118 \,(\mathrm{syst.})  \; \pm 0.012 \,(\mathrm{lumi.}) \; \mathrm{fb}$, per lepton flavour.
The inclusive $W^{\pm}Zjj$ production cross-section is measured differentially for several kinematic observables.
Finally, the measurements are used to constrain anomalous quartic gauge couplings by extracting $95$\% confidence level intervals on dimension-$8$ operators.
}

\AtlasRefCode{STDM-2018-35}

\PreprintIdNumber{CERN-EP-2024-082}

\arXivId{2403.15296}

\AtlasJournalRef{\JHEP 06 (2024) 192}
\AtlasDOI{10.1007/JHEP06(2024)192}

\hypersetup{pdftitle={ATLAS document},pdfauthor={The ATLAS Collaboration}}

\newcommand{\AtlasCoordFootnote}{
ATLAS uses a right-handed coordinate system with its origin at the nominal interaction point (IP)
in the centre of the detector and the \(z\)-axis along the beam pipe.
The \(x\)-axis points from the IP to the centre of the LHC ring,
and the \(y\)-axis points upwards.
Cylindrical coordinates \((r,\phi)\) are used in the transverse plane,
\(\phi\) being the azimuthal angle around the \(z\)-axis.
The pseudorapidity is defined in terms of the polar angle \(\theta\) as \(\eta = -\ln \tan(\theta/2)\).
Angular distance is measured in units of \(\Delta R \equiv \sqrt{(\Delta\eta)^{2} + (\Delta\phi)^{2}}\).}

\DeclareUnicodeCharacter{306}{\v{}}

\usepackage{rotating}

\begin{document}
 
\maketitle
 
\tableofcontents

\section{Introduction}
\label{sec:intro}

Vector boson scattering (VBS), $VV \to VV$ with $V= W/Z/\gamma$, is a key process with which to probe the $\text{SU}(2)_{\mathrm{L}} \times \text{U}(1)_{\mathrm{Y}}$ gauge symmetry of the electroweak (EW) theory  that determines the self-couplings of the vector bosons.
New phenomena beyond the Standard Model (SM) can alter the couplings of vector bosons, generating additional contributions to quartic gauge couplings compared with the SM predictions~\cite{Eboli:2003nq,Eboli:2006wa,Degrande:2012wf}.
 
\begin{figure}[t]
\begin{center}
\includegraphics[width=0.22\textwidth]{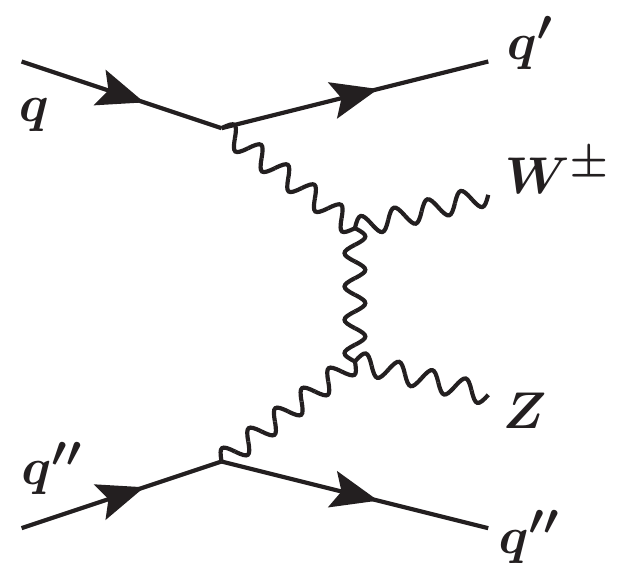} 
\hspace*{0.2cm}\includegraphics[width=0.22\textwidth]{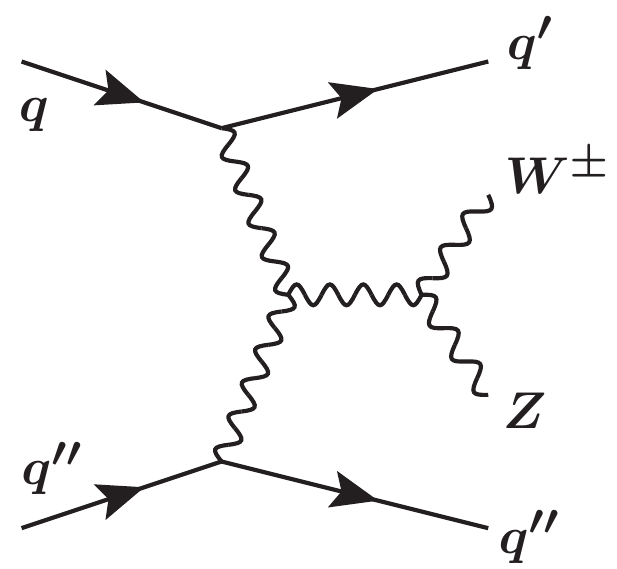} 
\hspace*{0.2cm}\includegraphics[width=0.22\textwidth]{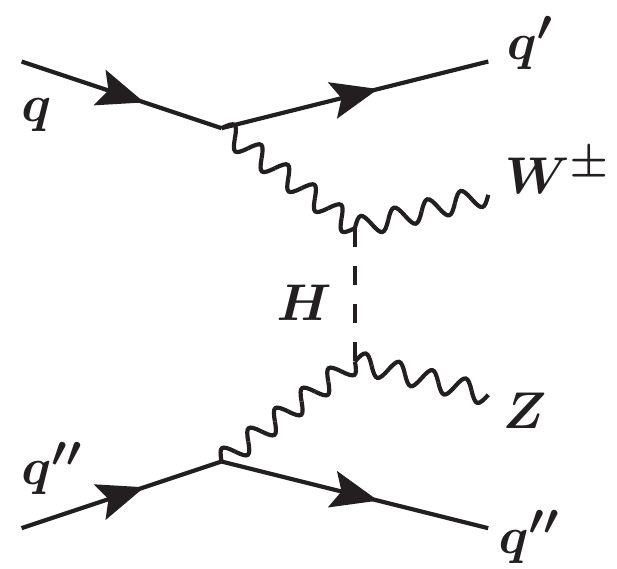} 
\hspace*{0.2cm}\hspace*{0.2cm}\includegraphics[width=0.22\textwidth]{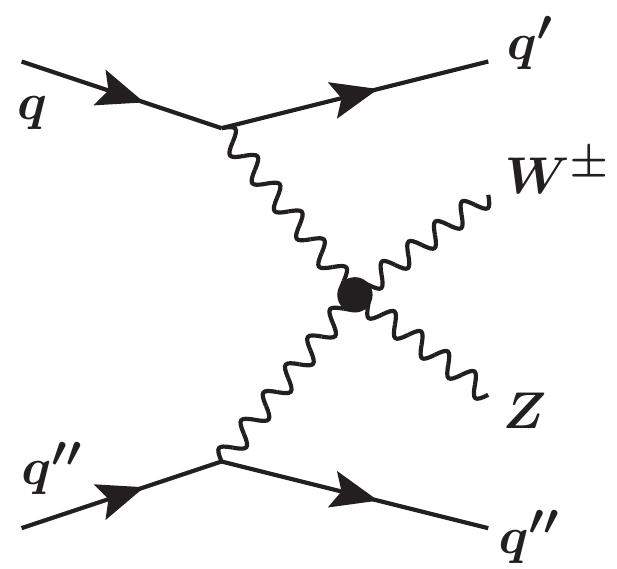} 
\caption{Representative diagrams at \LO of the \wzjjew production in $pp$ collisions.
}
\label{fig:diag_WZjjEW}
\end{center}
\end{figure}

\begin{figure}[t]
\begin{center}
\includegraphics[width=0.22\textwidth]{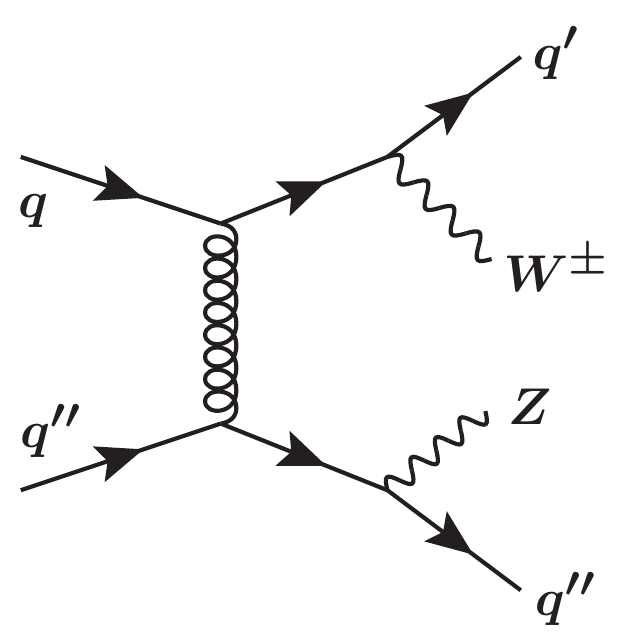} 
\hspace*{0.2cm}\includegraphics[width=0.22\textwidth]{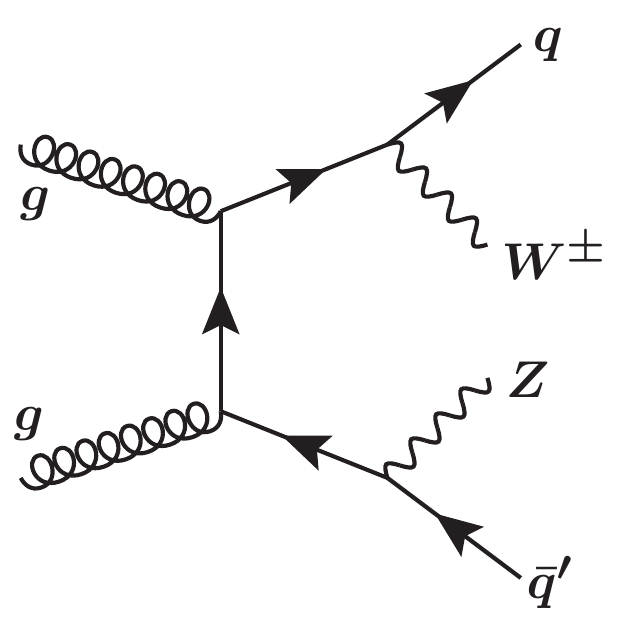} 
\hspace*{0.2cm}\includegraphics[width=0.22\textwidth]{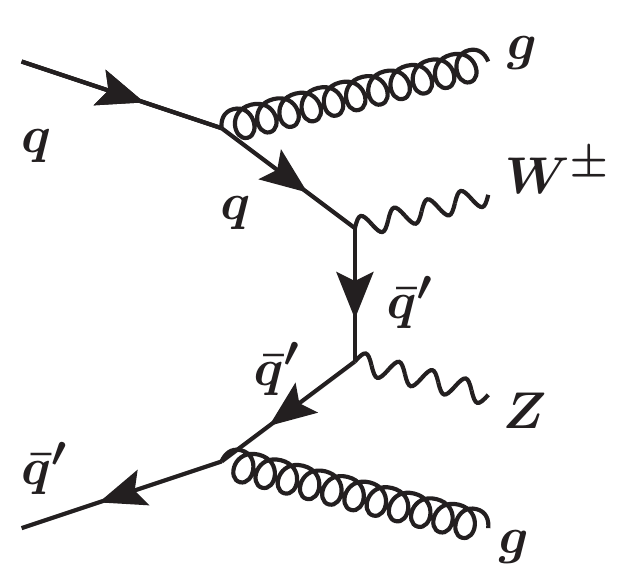} 
\caption{Representative diagrams at \LO of the \wzjjqcd production in $pp$ collisions.
}
\label{fig:diag_WZjjQCD}
\end{center}
\end{figure}

In proton--proton collisions, VBS is initiated by an interaction of two vector bosons radiated from the initial-state quarks, yielding a final state with two bosons and two jets, $VVjj$, in a purely electroweak process~\cite{Cahn:1986zv}.
VBS diagrams are not independently gauge invariant and cannot be studied separately from other processes leading to the same $VVjj$ final state~\cite{Accomando:2006mc}.
Two categories of processes give rise to $VVjj$ final states. The first category, which includes VBS contributions, involves exclusively weak interactions at Born level of order $\alpha_{\mathrm{EW}}^6$ including the boson decays, where $\alpha_{\mathrm{EW}}$ is the electroweak coupling constant.
It is referred to as EW production.
The second category involves both the quantum chromodynamics (QCD) and electroweak interactions at Born level of order $\alphas^2\alpha_{\mathrm{EW}}^4$, where $\alphas$ is the strong interaction coupling constant.
It is referred to as QCD production.
According to the SM a small interference occurs between electroweak and QCD production that also involves  both the strong and electroweak interactions at Born level of orders $\alphas\,\alpha_{\mathrm{EW}}^5$.
The sum of QCD and interference contributions are referred to as strong production, as opposite to the pure EW production.
Representative diagrams at \LO of \wzjjew and \wzjjqcd production are presented in Figures~\ref{fig:diag_WZjjEW} and \ref{fig:diag_WZjjQCD}, respectively.

A first observation of \wzjj electroweak production was reported by the ATLAS Collaboration~\cite{WZVBS36fb} using an integrated luminosity of $36$~$\ifb$ of $pp$ collision data collected at a centre-of-mass energy of $13$~\tev.
The observation was then confirmed by the CMS Collaboration using data corresponding to an integrated luminosity of $137$~$\ifb$~\cite{CMS:2020gfh}.
In these two results only integrated measurements of the electroweak \wzjj production cross-section were made, without further attempts to verify the modelling of key kinematic observables of the electroweak \wzjj processes by theory predictions.
The CMS and ATLAS Collaborations also used the measured \wzjj events to obtain constraints on the structure of quartic vector boson interactions in an Effective Field Theory (EFT) framework~\cite{CMS:2020gfh,ATL-PHYS-PUB-2023-002}.

This paper presents measurements of integrated and differential cross-sections for electroweak and inclusive \wzjj production, exploiting the fully leptonic decays of the $Z$ and $W$ bosons where final state charged leptons are electrons or muons.
The $pp$ collision data were collected with the ATLAS detector from $2015$ to $2018$ at a centre-of-mass energy of $\sqrt{s} = 13$~\TeV\ and correspond to an integrated luminosity of $140~\ifb$.
A multivariate discriminant constructed from a boosted decision tree (BDT) is used to separate electroweak and strong \wzjj production modes.
Their respective cross-sections are measured integrated over a fiducial phase space.
They are also measured differentially in three bins of the invariant mass of the two tagging jets, \mjj, and separately for events with exactly two jets or with more than two jets.
The cross-section of inclusive \wzjj production is also measured differentially as a function of various kinematic observables sensitive to the modelling from Monte Carlo (MC)  event generators or to new physics effects.
The distributions of the transverse mass of the \wz system, \mtwz, and of the BDT score are used in an EFT interpretation to constrain dimension-$8$ operators.


\section{ATLAS detector}
\label{sec:detector}


The ATLAS experiment~\cite{PERF-2007-01} at the Large Hadron Collider (LHC)~\cite{Evans:2008zzb} is a multipurpose particle detector
with a forward--backward symmetric cylindrical geometry and a near \(4\pi\) coverage in
solid angle.\footnote{\AtlasCoordFootnote}
It consists of an inner tracking detector (ID) surrounded by a thin superconducting solenoid
providing a \SI{2}{\tesla} axial magnetic field, electromagnetic and hadron calorimeters and a muon spectrometer (MS).
The inner tracking detector covers the pseudorapidity range \(|\eta| < 2.5\).
It consists of silicon pixel, silicon microstrip and transition radiation tracking detectors.
Lead/liquid-argon (LAr) sampling calorimeters provide electromagnetic (EM) energy measurements
with high granularity.
A steel/scintillator-tile hadron calorimeter covers the central pseudorapidity range (\(|\eta| < 1.7\)).
The endcap and forward regions are instrumented with LAr calorimeters
for both the EM and hadronic energy measurements up to \(|\eta| = 4.9\).
The muon spectrometer surrounds the calorimeters and is based on
three large superconducting air-core toroidal magnets with eight coils each.
The field integral of the toroids ranges between \num{2.0} and \SI{6.0}{\tesla\metre}
across most of the detector.
The muon spectrometer includes a system of precision tracking chambers and fast detectors for triggering.
A two-level trigger system is used to select events.
The first-level trigger is implemented in hardware and uses a subset of the detector information
to accept events at a rate below \SI{100}{\kHz}.
This is followed by a software-based trigger that
reduces the accepted event rate to \SI{1}{\kHz} on average
depending on the data-taking conditions.
An extensive software suite~\cite{ATL-SOFT-PUB-2021-001} is used in the reconstruction and analysis of real
and simulated data, in detector operations, and in the trigger and data acquisition systems of the experiment.


\section{Phase space for cross-section measurements}
\label{sec:phase_space}

The \wzjj cross-sections are measured in a fiducial phase space that is defined by the kinematics of the final-state leptons, electrons or muons, associated with the \Wpm and $Z$ boson decays.
Leptons produced in the decay of a hadron, a $\tau$-lepton, or their descendants are not considered in the definition of the fiducial phase space and are treated as background.
At particle level, the kinematics of the charged leptons after quantum electrodynamics final-state radiation are `dressed'
by including contributions from photons with an angular distance $\Delta R < 0.1$
from the lepton. Dressed charged leptons, and final-state neutrinos that do not originate from hadron or $\tau$-lepton decays, are matched to the \Wpm and $Z$ boson decay products using a MC generator-independent algorithmic approach, called the `resonant shape' algorithm~\cite{ATLASWZ_8TeV}.
This algorithm is  based on the value of an estimator expressing the product of the nominal line-shapes of the $W$ and $Z$ resonances,
\begin{equation*}\label{eq:Pk}
P = \left| \frac{1}{ m^2_{(\ell^+,\ell^-)} - \left(m_Z^{\textrm{PDG}}\right)^2 + \textrm{i} \; \Gamma_Z^{\textrm{PDG}} \; m_Z^{\textrm{PDG}} } \right|^2  \times \; \left| \frac{1}{ m^2_{(\ell',\nu_{\ell'})} - \left(m_W^{\textrm{PDG}}\right)^2 + \textrm{i} \; \Gamma_W^{\textrm{PDG}} \; m_W^{\textrm{PDG}} } \right|^2 \, ,
\end{equation*}
where $m_Z^{\textrm{PDG}}$ ($m_W^{\textrm{PDG}}$) and $\Gamma_Z^{\textrm{PDG}}$ ($\Gamma_W^{\textrm{PDG}}$)  are the world average mass and total width of the $Z$ ($W$) boson, respectively, as reported by the Particle Data Group~\cite{Tanabashi:2018oca}.
The input to the estimator is the invariant mass $m$ of all possible pairs ($\ell^+,\ell^-$) and ($\ell',\nu_{\ell'}$)
satisfying the fiducial selection requirements defined in the next paragraph.
The final choice of which leptons are assigned to the $W$ or $Z$ bosons corresponds to the configuration  exhibiting the largest value of the estimator.
Using this association algorithm, the gauge boson kinematics can be computed using the kinematics of the associated leptons independently of any internal MC generator details.

The fiducial phase space of the measurement is the same as used in Ref.~\cite{WZVBS36fb}. 
It is defined at particle level by the following requirements:
the charged leptons from the $Z$ boson decay are required to have transverse momentum $\pt>15$~\GeV, the charged lepton from the \Wpm decay is required to have transverse momentum $\pt^\ell>20$~\GeV, the charged leptons from the \Wpm and $Z$ bosons are required to have $\abseta<2.5$ and the invariant mass of the two leptons from the $Z$ boson decay must be within $10$~\GeV{} of the nominal $Z$ boson mass, taken from the world average value, $m_Z^{\textrm{PDG}} = 91.1876$~\GeV~\cite{Tanabashi:2018oca}.
The $W$ boson transverse mass, defined as $m_{\mathrm{T}}^{W} = \sqrt{2 \cdot p_{\mathrm{T}}^\nu \cdot p_{\mathrm{T}}^\ell \cdot[1 -\cos{\Delta \phi(\ell, \nu)}]}$, where $\Delta \phi(\ell, \nu)$
is the angle between the lepton and the neutrino in the transverse plane and $\pt^\nu$ is the transverse momentum of the neutrino, is required to be $m_{\mathrm{T}}^{W}>30$~\GeV.
The angular distance between the charged lepton from the $W^\pm$ decay and each of the charged leptons from the $Z$ decay is required to be  $\Delta R>0.3$, and the angular distance between the two leptons from the $Z$ decay is required to be $\Delta R>0.2$.
 
In addition to these requirements that define an inclusive phase space, at least two jets with $\pt>40$~\GeV\ and $|\eta_j|< 4.5$ are required.
These particle-level jets are reconstructed from stable particles with a lifetime of $\tau>30$~ps in the simulation after parton showering, hadronisation, and decay of particles with $\tau<30$~ps.
Muons, electrons, neutrinos and photons associated with $W$ and $Z$ decays are excluded.
The particle-level jets are reconstructed using the anti-$k_{t}$~\cite{antikt,Fastjet} algorithm with a radius parameter $R$ = $0.4$.
The angular distance between all selected leptons and jets is required to be $\Delta R (j,\ell)>0.3$.
If this requirement is not satisfied, the jet is discarded.
The invariant mass, $m_{jj}$, of the two highest-\pT jets in opposite hemispheres, $\eta_{j1} \cdot \eta_{j2} < 0$, is required to be $m_{jj}>500$~\GeV\ to enhance the sensitivity to VBS processes.
These two jets are referred to as tagging jets.
Finally, hard processes with a $b$-quark in the initial or final states at the matrix-element level, such as $tZj$ production,  are not considered as part of the \wzjjew events.
The production of $tZj$ results from a $t$-channel exchange of a $W$ boson  between a $b$- and a $u$-quark giving a final state with a $t$-quark, a $Z$ boson and a light-quark jet, but does not include diagrams with gauge boson couplings.


\section{Signal and background simulation}
\label{sec:samples}

 
The \MGNLO[2.6.5]~\cite{Alwall:2014hca} MC event generator, interfaced with \PYTHIA[8.240]~\cite{Sjostrand:2014zea} using the A14 tune~\cite{ATL-PHYS-PUB-2014-021} for the modelling of the parton shower (PS), hadronisation and underlying event, was used to model \wzjj\ events.
The \NNPDF[3.0nlo]~\cite{Ball:2014uwa} parton distribution function (PDF) set was used for the hard-scattering process,  while the \NNPDF[2.3lo]~\cite{Ball:2014uwa}  PDF set was used for the PS.
The dipole recoil scheme~\cite{Cabouat:2017rzi} is used for the PS.
The default dynamic renormalisation and factorisation scales set by \MGNLO~\cite{Hirschi:2015iia} were used.
A first MC event sample, referred to as \wzjjew, includes processes of order six (zero) in $\alpha_\mathrm{EW}$ ($\alphas$).
In this sample, which includes  VBS diagrams, two additional jets originating from electroweak vertices from matrix-element partons are included in the final state.
Diagrams with a $b$-quark in either the initial or final state, i.e.\ $b$-quarks in the matrix-element calculation, are not considered.
This sample provides a \LO prediction for the \wzjjew signal process.
A second MC event sample, referred to as \wzjjqcd, includes processes of order four in $\alpha_\mathrm{EW}$ in the matrix-element.
Matrix elements containing three leptons, one neutrino and up to two jets in the final state were calculated at \NLO QCD and merged with the PS from  \PYTHIA[8.210] using the \FXFX scheme~\cite{Frederix:2012ps}.
This \wzjjqcd sample includes matrix-element $b$-quarks.
Interferences between the \wzjjew and \wzjjqcd processes, labelled \wzjjint,  include only contributions to the squared matrix-element of order one in $\alphas$.
Their contribution is simulated at \LO in a third MC sample using \MGNLO[2.6.5]~\cite{Alwall:2014hca} and the same parameters as used for \wzjjew events.
The contribution of interferences is found to be positive and approximately $6\%$ of the \wzjjew cross-section in the fiducial phase space at particle level and maximaly $12\%$ of the \wzjjew cross-section for events with three or more jets of $\pt > 25$~\GeV.

For the estimate of the PS uncertainties in the simulation of the \wzjjew process, an alternative MC sample was simulated using \MGNLO[2.6.5] interfaced with \HERWIG[7.1]~\cite{Bellm:2015jjp,Bellm:2017bvx}.
For the estimate of modelling uncertainties and comparisons to data, alternative MC samples of \wzjjqcd and \wzjjew processes were generated with the \SHERPA[2.2.12]~\cite{Bothmann:2019yzt}   MC event generator.
For \wzjjqcd processes, the matrix elements contain all diagrams with four electroweak vertices.
They were calculated for up to one parton at next-to-leading-order (NLO) in QCD and up to three partons at leading-order (LO) using Comix~\cite{Gleisberg:2008fv} and \OPENLOOPS~\cite{Cascioli:2011va}, and merged with the \Sherpa PS~\cite{Schumann:2007mg} according to the ME+PS$@$NLO prescription~\cite{Hoeche:2012yf}.
The \NNPDF[3.0nnlo] PDF set was used along with the associated set of tuned parton-shower parameters developed by the \Sherpa authors.
For \wzjjew production, processes of order six (zero) in $\alpha_\mathrm{EW}$ ($\alphas$) are included.
Two additional jets originating from electroweak vertices from matrix-element partons are included in
the final state.
Diagrams with a $b$-quark in either the initial or final state are not considered.
The LO-accurate matrix elements for up to one additional parton were matched to a PS based on Catani--Seymour dipole factorisation~\cite{Gleisberg:2008fv,Schumann:2007mg} using the MEPS@LO prescription~\cite{Hoeche:2011fd,Hoeche:2012yf,Catani:2001cc,Hoeche:2009rj}.
The contribution of interferences is not simulated using \SHERPA[2.2.12] but the \wzjjint prediction from  \MGNLO[2.6.5] is added to the \wzjjew and \wzjjqcd \SHERPA[2.2.12] predictions to form the \wzjj \SHERPA[2.2.12] prediction.

The background sources include processes with two or more electroweak gauge bosons, namely $ZZ$, $WW$ and $VVV$ ($V=W,Z$); processes with top quarks, such as \tt and \ttV, single top and \tzj; and processes with gauge bosons associated with jets or photons ($Z+j$ and $Z\gamma$).
MC simulation is used to estimate the contribution from  background processes with three or more prompt leptons.
Background processes with at least one misidentified lepton are evaluated using data-driven techniques and simulated events are used to assess the systematic uncertainties of these backgrounds.

The \SHERPA[2.2.2] event generator was used to simulate both the $q\bar{q}$ and $gg$-initiated $ZZ$ processes, including $H \rightarrow ZZ$ production, using the \NNPDF[3.0nnlo] PDF set.
It provides a matrix element calculation accurate to NLO in $\alpha_{\mathrm{s}}$ for 0- and 1-jet final states, and LO accuracy for 2- and 3-jet final states.
The \zzjjew production was modelled using the \MGNLO[2.6.7] generator with matrix elements calculated at \LO in QCD and with the \NNPDF[3.0nlo] PDF set.
The events were interfaced with \PYTHIA[8.230] using the A14 tune and the \NNPDF[2.3lo] PDF set.
 
The production of \ttV\ events was modelled using the \MGNLO[2.3.3]~\cite{Alwall:2014hca} generator at \NLO with the \NNPDF[3.0nlo] PDF\@. The events were interfaced to \PYTHIA[8.210] using the A14 tune and the \NNPDF[2.3lo] PDF set.
The \tzj process was modelled using the \MGNLO[2.3.3]~\cite{Alwall:2014hca} generator at \NLO with the \NNPDF[3.0nlo] PDF. The events were interfaced with \PYTHIA[8.230] using the A14 tune and the \NNPDF[2.3lo] PDF set.
Finally, the on-shell production of triboson events with fully leptonic decays were simulated by the \SHERPA[2.2.2] event generator at NLO accuracy with zero additional partons and at LO accuracy with one and two additional partons and using the \NNPDF[3.0nnlo]  PDF set.
The decays of bottom and charm hadrons were simulated using the \EVTGEN[1.2.0] program~\cite{Lange:2001uf}, except for processes modelled using \SHERPA.

All generated MC events were passed through the ATLAS detector simulation~\cite{Aad:2010ah}, based on
\GEANT~\cite{Agostinelli:2002hh}, and processed using the same reconstruction software as used for the data.
The event samples include the simulation of additional $pp$ interactions (pile-up)
generated with \PYTHIA~8.186 using the \NNPDF[2.3lo] PDF set
and the A3~\cite{ATL-PHYS-PUB-2016-017} set of tuned parameters. 
Simulated events were reweighted to match the pile-up conditions observed in the data.
Furthermore, the lepton and jet momentum scale and resolution, the lepton reconstruction, identification, isolation and trigger efficiencies and the pile-up jets and $b$-jet veto efficiencies in the simulation were corrected to match those measured in data.


\section{Event selection}
\label{sec:selection}


Only data recorded with stable beam conditions and with all relevant detector subsystems operational are considered.
Candidate events are selected using triggers requiring at least one electron or muon~\cite{TRIG-2016-01,ATLAS:2019dpa,ATLAS:2020gty}.
These triggers require leptons to satisfy different transverse momentum thresholds and isolation criteria, which depend on the data-taking run period and the instantaneous luminosity.
The combined efficiency of these triggers is $99\%$ for \wzjj events satisfying the offline selection criteria.
Events are required to have a primary vertex compatible with the LHC luminous region inside the ATLAS detector.
The primary vertex is defined as the reconstructed vertex with at least two charged-particle tracks that has the largest sum of the $\pT^{2}$ for the associated tracks.

Muon candidates are identified by tracks reconstructed in the MS and matched to tracks
reconstructed in the ID\@. Muons are required to satisfy a `medium' identification selection~\cite{ATLAS:2016lqx,ATLAS:2020auj}.
The efficiency of this selection averaged over \pt and $\eta$ is larger than $98\%$.
The muon momentum is measured by combining the MS measurement, corrected for the energy deposited in
the calorimeters, and the ID measurement.
The $\pt$ of the muon must be greater than $15$~\GeV{} and its pseudorapidity must satisfy $|\eta|<2.5$.
 
Electron candidates are reconstructed from energy clusters in the electromagnetic calorimeter matched
to ID tracks. Electrons are identified using a discriminant that is the value of a likelihood
function constructed with information about the shape of the electromagnetic showers in the calorimeter,
the track properties and the quality of the  track-to-cluster matching for the candidate~\cite{ATLAS:2019qmc}.
Electrons must satisfy a `medium' likelihood requirement, which provides an overall identification efficiency of $90\%$.
The electron momentum is computed from the cluster energy and the direction of the track. The $\pt$ of
the electron must be greater than $15$~\GeV{} and the pseudorapidity of the cluster must satisfy
$\abseta< 1.37$ or $1.52 < \abseta < 2.47$.
 
Electron and muon candidates are required to originate from the primary vertex. Thus, the significance of
the track's transverse impact parameter calculated relative to the beam line, $|{d_{0}/\sigma_{d_{0}}}|$,
must be smaller than $3.0$ for muons and less than $5.0$ for electrons. Furthermore, the longitudinal impact parameter,
$z_{0}$ (the difference between the value of $z$ of the point on the track at which $d_{0}$ is defined and the
longitudinal position of the primary vertex), is required to satisfy $|z_0\cdot \sin(\theta)|<0.5$~mm.

Electrons and muons are required to be isolated from other particles using both calorimeter-cluster and ID-track information.
The isolation requirement for electrons is tuned for an efficiency of  at least $88\%$ for $\pT>15$~\GeV\ and
at least $99\%$ for $\pT>60$~\GeV~\cite{ATLAS:2019qmc}, while particle-flow-based isolation variables are used for muons, providing an efficiency above $90\%$ for $\pT>15$~\GeV\ and at least $99\%$ for $\pT>60$~\GeV~\cite{ATLAS:2020auj}.

Jets of hadrons are reconstructed using a particle-flow algorithm based on noise-suppressed positive-energy topological clusters in the calorimeters~\cite{ATLAS:2017ghe}.
Energy deposited in the calorimeters by charged particles is subtracted and replaced by the momenta of tracks which are matched to those topological clusters.
Tracks which are not matched to the primary vertex are not used in jet reconstruction, which effectively removes contribution from charged particle pile-up.
The jets are clustered using the anti-$k_{t}$ algorithm with a radius parameter $R = 0.4$.
They are  calibrated according to in situ measurements of the jet energy scale~\cite{ATLAS:2020cli}.
All jets must have $\pt>25$~\GeV{} and be reconstructed in the pseudorapidity range $|\eta|<4.5$.
To reduce the impact of jets originating from pile-up interactions, jets with
$|\eta|<2.4$ and $\pt < 60$~\GeV, or with $2.5 < |\eta| < 4.5$ and $\pt < 120$~\GeV, are required to satisfy the `tight' and `loose' working points of the jet vertex~\cite{Aad:2015ina} and forward jet vertex~\cite{ATL-PHYS-PUB-2019-026} tagging algorithms, respectively.
Jets with $|\eta|<2.5$ containing a $b$-hadron are identified with a deep-learning neural network (NN)~\cite{FTAG-2019-07} which uses distinctive features of $b$-hadron decays in terms of the impact parameters of the tracks and the displaced vertices reconstructed in the inner detector.
Jets initiated by $b$-quarks are selected by setting the algorithm's output threshold such that an $70\%$ $b$-jet selection efficiency is achieved in simulated $t\bar{t}$ events, with a rejection factor of $40$ against light-flavour jets.
Corrections to the flavour-tagging efficiencies are based on data-driven calibration analyses~\cite{FTAG-2019-07}.

The transverse momentum of the neutrino is estimated from the missing transverse momentum in the event, \met,
calculated as the negative vector sum of the transverse momentum of all identified hard physics objects
(electrons, muons, jets), with a contribution from an additional soft term.
This soft term is calculated from ID tracks matched to the primary vertex and not assigned to any of the hard objects~\cite{ATLAS-CONF-2018-023}.

Events are required to contain exactly three lepton candidates satisfying the selection criteria described above.
To ensure that the trigger efficiency is well determined, at least one of the candidate leptons is required to have
$\pt > 25$~\GeV\ for $2015$ and $\pt > 27$~\GeV\ for $2016$--$2018$ data, as well as being  geometrically matched to a lepton that was selected by the trigger.
This detector-level requirement that the transverse momentum of the leading lepton be above  $27$~\GeV\ is not present in the definition of the fiducial phase space.
It would reduce the acceptance of the fiducial phase space by only $0.02\%$.

To suppress background processes with at least four prompt leptons, events with a fourth lepton candidate
satisfying looser selection criteria are rejected. For this looser selection, the lepton $\pt$ requirement is
lowered to $\pt>5$~\GeV{}, electrons are allowed to be reconstructed in the region $1.37< \abseta< 1.52$ and `loose' identification requirements~\cite{ATLAS:2019qmc,ATLAS:2020auj} are used for both the electrons and muons.
A less stringent requirement is applied for electron isolation and is based only on ID track information.
No dedicated identification algorithm is used to suppress events with electrons and muons originating from the decay of $\tau$-leptons.
 
Candidate events are required to have at least one pair of leptons with the same flavour and opposite charge,
with an invariant mass that is consistent with $m_Z^{\textrm{PDG}}$ to within $10$~\gev.
This pair is considered to be the $Z$ boson candidate. If more than one pair can be formed, the pair whose invariant mass is closest to $m_Z^{\textrm{PDG}}$ is taken as the \Zboson\ boson candidate.
The remaining third lepton is assigned to the $W$ boson decay\footnote{For true \wzjj events with same flavour decays, the probability to mis-assign a lepton with its mother boson is of $2.5\%$.}.
The transverse mass of the $W$ candidate, computed using
\met and the \pt of the associated lepton, is required to be greater than $30$~\GeV.
 
Backgrounds originating from misidentified leptons are suppressed by requiring the lepton associated with the $W$ boson to satisfy more stringent selection criteria. The transverse momentum of this lepton is therefore required to be greater than $20$~\GeV. This lepton is also required to satisfy the `tight'  identification and isolation requirements~\cite{ATLAS:2019qmc,ATLAS:2020auj}, which results in an efficiency between $90\%$ and $98\%$ depending on \pt for muons and an overall efficiency of $85\%$ for electrons.

To select \wzjj\ candidates, events are further required to be associated with at least two `tagging' jets.
The leading tagging jet is selected as the highest-\pt\ jet in the event with $\pt > 40$~\GeV.
The second tagging jet is selected as the one with the highest \pt\ among the remaining jets that have a pseudorapidity of opposite sign to the first tagging jet and a $\pt > 40$~\GeV.
These two jets are required to satisfy $m_{jj} > 150$~\GeV, in order to minimise the contamination from triboson processes.

The final signal region (SR) for VBS processes is defined by requiring that the invariant mass of the two tagging jets, $m_{jj}$, be greater than $500$~\GeV\ and that no $b$-jet be present in the event.


\section{Background estimation}
\label{sec:background}

The background sources are classified into two groups:  events where all candidates are prompt leptons or are produced in the decay of a $\tau$-lepton (irreducible background) and events where at least one of the candidate leptons is not a prompt lepton (reducible background). Candidates that are not prompt leptons are also called  `misidentified' or `fake' leptons.

The main sources of irreducible background arise from $ZZ$ and $t\bar{t} + V$ (where $V$ = $Z$ or $W$).
These irreducible backgrounds are modelled using MC simulations.
Data in two dedicated control regions (CR), referred to as $ZZ$-CR and $b$-CR, respectively, are used to constrain the normalisations of the \zzjjqcd\ and $t\bar{t} + V$ backgrounds.
The control region $ZZ$-CR, enriched in $ZZ$ events, is defined by applying the \wzjj event selection defined in Section~\ref{sec:selection}, but instead of vetoing a fourth lepton, events must have at least a fourth lepton candidate with looser identification requirements.
This region is dominated by \zzjjqcd events with a small contribution of \zzjjew\ events.
The control region $b$-CR, enriched in \ttV and \tzj events, is defined by selecting \wzjj\ candidate events having at least one reconstructed $b$-jet.
The remaining sources of irreducible background are \zzjjew and $VVV$ events.
Their contributions in the control and signal regions are estimated from MC simulations.

The reducible backgrounds originate from $Z+j$, $Z\gamma$, $t\bar{t}$, $Wt$ and $WW$ production processes.
These backgrounds are estimated by using a data-driven method based on the inversion of a global matrix containing the efficiencies and the misidentification probabilities for prompt and fake leptons~\cite{ATLASWZ_8TeV,Aad:2014pda,WZPol36fb,ATLAS:2022swp}.
The method exploits  the classification of the leptons as  loose  or tight  candidates and  the probability that a fake lepton is misidentified as a loose or tight lepton candidate.
Tight lepton candidates are signal lepton candidates as defined in Section~\ref{sec:selection}.
Loose lepton candidates are leptons that do not meet the isolation and identification criteria of signal lepton candidates but satisfy only looser criteria.
The misidentification probabilities for misidentified leptons are determined from data using dedicated control samples each enriched  in misidentified leptons from light- or heavy-flavour jets and from photon conversions, respectively.
The lepton efficiencies for prompt leptons are determined as detailed in Refs.~\cite{ATLAS:2019qmc,ATLAS:2016lqx,ATLAS:2020auj}.
The lepton efficiencies and misidentification probabilities are parameterised as a function of the \pt and $\eta$ of leptons and combined with event rates in data samples of \wzjj candidate events where at least one and up to three of the leptons are loose.
Then, solving a system of linear equations, the number of events with at least one misidentified lepton  is obtained, which represents the amount of reducible background in the \wzjj sample.
The method allows the shape of any kinematic distribution of reducible background events to be estimated.
Another independent method of assessing the reducible background was also considered.
This method estimates the amount of reducible background using MC simulations scaled to data by process-dependent factors determined from the data-to-MC comparison in dedicated control regions.
Agreement within $20\%$ with the matrix method estimate is obtained in both yield and shape of the distributions of irreducible background events.
This level of agreement is compatible with the systematic uncertainties determined for the matrix method estimate.

The number of events expected from MC simulation and data-driven background estimates are summarised in Table~\ref{tab:VBSyield_prefit} and compared to the number of observed data events, for the signal region and the two control regions.
All sources of uncertainties, as described in Section~\ref{sec:systematics}, are included.
The expected contribution of \wzjjew events in the \wzjj signal region is about $16\%$ while $64\%$ of the events are expected to arise from \wzjjqcd production.
The contributions of \wzjj events with $\tau$-lepton decays, amounting to $5.3\%$ of all \wzjj events, are included in the \wzjjew, \wzjjqcd and \wzjjint contributions.

\begin{table}[t] 
\caption{Expected and observed numbers of events in the \wzjj signal region and in the two control regions, before the fit described in Section~\ref{sec:meth_inclusive}.
The expected number of \wzjjew, \wzjjqcd and \wzjjint events and the estimated number of background events from the other processes are shown.
The sum of the backgrounds containing misidentified leptons is labelled `Misid. leptons'.
The total systematic uncertainties are quoted.
\label{tab:VBSyield_prefit}
}
\centering

\begin{tabular}{l r@{\,}l!{$\pm$}r@{\,}l r@{\,}l!{$\pm$}r@{\,}l r@{\,}l!{$\pm$}r@{\,}l r@{\,}l!{$\pm$}r@{\,}l }
\toprule
& \multicolumn{4}{c}{SR, $\Njets = 2$}  & \multicolumn{4}{c}{SR, $\Njets \ge 3$}  & \multicolumn{4}{c}{$b$-CR} & \multicolumn{4}{c}{$ZZ$-CR} \\
\midrule
Data & \multicolumn{4}{c}{169}  & \multicolumn{4}{c}{477}  &   \multicolumn{4}{c}{666} & \multicolumn{4}{c}{210}  \\
Total pred. & 231& & 12&  & 550& & 50&  &   660& & 40& & 205& & 11&  \\
\midrule
$WZjj$-EW & 65&.0 & 3&.5  & 60& & 6&  &   4&.82 & 0&.28 & 0&.725 & 0&.014  \\
$WZjj$-QCD & 125& & 9&  & 380& & 50&  &   77& & 18& & 6&.2 & 0&.7  \\
$WZjj$-INT & 1&.3 & 0&.6  & 5&.3 & 2&.6  &   0&.58 & 0&.29 & 0&.22 & 0&.11  \\
$t\bar{t}+V$ & 0&.66 & 0&.04  & 20&.2 & 0&.7  &   289& & 10& & 9&.89 & 0&.28  \\
$tZj$ & 8&.78 & 0&.34  & 19&.7 & 1&.2  &   134& & 4& & 0&.432 & 0&.005  \\
$ZZ$-QCD & 9&.6 & 0&.4  & 32&.0 & 2&.5  &   10&.1 & 0&.6 & 159& & 9&  \\
$ZZ$-EW & 2&.2 & 0&.6  & 4&.4 & 1&.1  &   0&.25 & 0&.06 & 23& & 6&  \\
$VVV$ & 0&.41 & 0&.10  & 2&.0 & 0&.5  &   0&.39 & 0&.10 & 4&.1 & 1&.1  \\
Misid. leptons & 18& & 4&  & 28& & 7&  &   150& & 40& & 1&.7 & 0&.5  \\
\bottomrule
\end{tabular}


\end{table}


\section{Measurements methodology}
\label{sec:analysis}

 
\subsection{\wzjjew and \wzjjnew integrated measurements}\label{sec:meth_inclusive}

Events associated to \wzjjew production are separated from other \wzjj events originating from \wzjjqcd production and from the interference \wzjjint, based on their kinematic properties as modelled by MC predictions.
The production cross-section of the two populations of events, \sigew and \sigqcd, are then measured simultaneously.
 
Given the small contribution to the signal region of \wzjjew processes, a multivariate discriminant is used to separate the signal from the backgrounds.
A BDT, as implemented in the TMVA package~\cite{Hocker:2007ht}, is used to exploit the kinematic differences between the \wzjjew signal and the \wzjjqcd and other backgrounds.
The BDT is trained and optimised on simulated events from the SR to separate \wzjjew events from all other processes.
 
Similarly to Ref.~\cite{WZVBS36fb}, $15$ variables are used as inputs to the BDT discriminant. 
The variables can be classified into three categories: jet-kinematics variables, vector-bosons-kinematics variables, and variables related to both jets and leptons kinematics.
The variables related to the kinematic properties of the two tagging jets are the invariant mass of the two jets, $m_{jj}$, the transverse momenta of the jets, \ptji and \ptjii, the difference in pseudorapidity and azimuthal angle between the two jets, $\Delta \eta_{jj}$ and $\Delta \phi_{jj}$, the rapidity of the leading jet and the jet multiplicity.
The variables related to the kinematic properties of the vector bosons are the transverse momenta of the $W$ and $Z$ bosons, the pseudorapidity of the $W$ boson, the absolute difference between the rapidities of the $Z$ boson and the lepton from the decay of the $W$ boson, $|y_Z - y_{\ell,W}|$, and the transverse mass of the \wz system \mtwz.
The pseudorapidity of the $W$ boson is reconstructed using an estimate of the longitudinal momentum of the neutrino obtained using the neural-network based method developed in Ref.~\cite{WZPol36fb}.
The \mtwz observable is reconstructed following Ref.~\cite{ATLASWZ_8TeV}.
The variables that relate the kinematic properties of jets and leptons are the distance in the pseudorapidity--azimuth plane between the $Z$ boson and the leading jet, $\Delta R(j_1,Z)$,
the event balance $R_{p_\mathrm{T}}^{\mathrm{hard}}$, defined as the transverse component of the vector sum of the $WZ$ bosons and tagging jets momenta, normalised to their scalar \pt sum, and, finally the centrality of the $WZ$ system relative to the tagging jets, defined as $\zeta_{\mathrm{lep.}} = \text{min}(\Delta\eta_{-},\Delta\eta_{+})$, with $\Delta\eta_{-} = \text{min}(\eta_{\ell}^{W},\eta_{\ell_2}^{Z},\eta_{\ell_1}^{Z}) - \text{min}(\eta_{j_1},\eta_{j_2})$ and $\Delta\eta_{+} = \text{max}(\eta_{j_1},\eta_{j_2}) - \text{max}(\eta_{\ell}^{W},\eta_{\ell_2}^{Z},\eta_{\ell_1}^{Z})$, where $\eta_{\ell_1}^{Z}$ and $\eta_{\ell_2}^{Z}$ are the pseudorapidity of the leading and sub-leading leptons associated to the $Z$ boson.
The good modelling by MC simulations of the distribution shapes and the correlations of all input variables to the BDT is verified in a validation region defined by requiring $150 < \mjj < 500$~\GeV.
The modelling of input variables to the BDT, and in particular of the \mjj distribution, is also verified in a sub-signal region enriched in \wzjjqcd events. This sub-signal region is constructed using an Adversarial Neural Network (ANN) developed to separate \wzjjqcd from \wzjjew events in the SR, without introducing a bias in the shape of the \mjj distribution.
The output score of this ANN is used to select a sub-sample of events in the SR enriched in \wzjjqcd production. For these events, a good modelling of the shapes of the \mjj distribution and of the other BDT input observables is observed.
Finally the good modelling by MC simulations of the kinematic of \wzjj events is verified independently by the measurement of differential cross-sections of key observables as presented in Section~\ref{sec:wzjj_diff_method} and whose results are shown in Section~\ref{sec:wzjj_diff_results}.

To better separate the \ttV and \tzj events in the \bCR and to constrain the normalisation of both contributions using data in this CR, a second BDT discriminant, labelled \CRbBDT, is built using $16$ input variables.
The input variables used are related to the kinematics of tagging jets: \ptji, \etaji, \phiji, \ptjii, \dEtajj and \dPhijj; to the kinematics of vector bosons: \ptz, \etaz, \phiz, \ptw, \mwz, \dEtaWZ and \dPhiWZ; to the jet mutiplicites: \Njets, \Nbjets; and to the invariant mass of final state particles: \mllljets, defined as the invariant mass of the three charged leptons and all jets with a $\pt > 25$~\GeV.
The \CRbBDT is trained and optimised on simulated events to separate \tzj from \ttV events.
Its discriminant output is labelled \CRbBDT score.

In the \wzjj signal region, the distribution of the BDT score is used in a maximum-likelihood fit to separate \wzjjew events from \wzjjnew events and to measure their respective fiducial cross-sections.
To minimise the dependency of the measurement on possible mis-modelling of \wzjjqcd and \wzjjew events by the MC generators, the \wzjj SR is divided in two categories corresponding to events with exactly two jets of $\pt > 25$~\GeV or with three or more jets.
An extended likelihood is built from the product of four likelihoods corresponding to the BDT score distributions in the two event categories of the \wzjj SR, the \CRbBDT distribution in the \bCR and the $m_{jj}$ distribution in the \ZZCR.
The inclusion of the two control regions in the fit allows the yields of the \ttV, \tzj and \zzjjqcd backgrounds to be constrained by data.
The shapes of these backgrounds are taken from MC predictions and can vary within the uncertainties affecting the measurement as described in Section~\ref{sec:systematics}.
The normalisations of these backgrounds are introduced in the likelihood as parameters, labelled  \muttV, \mutZj and \muZZ for \ttV, \tzj and \zzjjqcd backgrounds, respectively.
They are treated as unconstrained nuisance parameters that are determined mainly by the data in the respective control region.
The normalisation and shape of the other irreducible backgrounds are taken from MC simulations and are allowed to vary within their respective uncertainties.
The distribution of the reducible background is estimated from data using the matrix method presented in Section~\ref{sec:background} and is allowed to vary within its uncertainty.
Each source of systematic uncertainty is implemented in the likelihood function as a nuisance parameter with a Gaussian constraint.

The fiducial cross-sections \sigew and \sigqcd integrated over the two event categories in \njets are determined as parameters of a binned fit to data.
The fit parameters are directly linked to normalisation factors of the \wzjjew, \wzjjqcd and \wzjjint MC templates as
 
\begin{equation}
\begin{array}{l l l}
\sigew & = & \displaystyle \sum_{i=1}^{2} \; \muewi \cdot \sigewthi \; , \\
\sigqcd & = & \displaystyle \sum_{i=1}^{2} \; \left( \muqcdi \cdot \sigqcdthi + \muinti \cdot \sigintthi \right) \; , \\
& = & \displaystyle \sum_{i=1}^{2} \;\left(  \muqcdi \cdot \sigqcdthi + \sqrt{\muewi} \cdot \sqrt{\muqcdi} \cdot \sigintthi  \right) \; ,
\end{array}
\end{equation}

where $i$ runs over the two categories of the SR with exactly two jets of  $\pt > 25$~\GeV or with three or more jets, \sigewthi, \sigqcdthi and \sigintthi are the theoretical cross-sections predicted by the MC for the respective processes in each category, and \muewi and \muqcdi are the normalisation factors of the \wzjjew and \wzjjqcd MC templates in each category.
The contributions from $\tau$-decay leptons are included in the \wzjjew, \wzjjqcd and \wzjjint MC templates, assuming that measured cross-sections are flavour independent.
By convention, the measured fiducial cross-sections correspond to a given decay channel \wz\ $\rightarrow \ell^{'\pm} \nu \ell^+ \ell^-$, where $\ell$ and $\ell^{'}$ are either an electron or a muon.
The contribution of the interference between \wzjjew and \wzjjqcd production processes, \wzjjint, is introduced in the fit using a separate MC template.
Its contribution is estimated to be very low at detector level, of the order of $5.3\%$ of the \wzjjew contribution  as predicted by \MGPy8.
The kinematics and shape of the \BDTscore distribution of \wzjjint events is very close to the one of \wzjjqcd events.
Therefore, with the statistical sensitivity of the present data sample, this contribution can not be extracted alone from data.
The normalisation of the interference contribution, \muinti, is not fixed in the fit to the predicted \MGPy8 cross-section but is linked to the measured \sigew and \sigqcd cross-section as $\muinti  =  \sqrt{\muewi} \cdot \sqrt{\muqcdi}$, and the \wzjjint contribution is included into the measured \sigqcd cross-section.

\subsection{\wzjjew and \wzjjnew differential measurements}

Events in the SR are separated in bins as a function of \njets, the exclusive multiplicity of jets with a $\pt > 25~\GeV$ or as a function of \mjj, the invariant mass of the two tagging jets.
Two bins in \njets, separating events with $\njets = 2$ and $\njets \ge 3$ and three bins in \mjj are used.
In each of the bins, the BDT score distribution is used to separate \wzjjew events and \wzjjnew events and therefore measure differentially the separate \sigew and \sigqcd production cross-sections.
A simultaneous fit to data of the BDT score distributions of events in each bin is performed to measure the fiducial cross-section in each bin.
For e.g. \wzjjew events, the fiducial cross-section is defined in each bin $i$ as
 
\begin{equation}
\sigewi = \muewi \cdot \sigewthi = \frac{N_{\mathrm{fit}}^i}{\mathcal{L} \cdot C_i} \, , \ \ \ C_i = \frac{N^i_{\mathrm{MC,\, det.}} }{N^i_{\mathrm{MC,\, part.}}} \, ,
\end{equation}

where \muewi is the normalisation factor of the \wzjjew MC template in bin $i$,  $N_{\mathrm{fit}}^i$ is the number of extracted \wzjjew events in data, $\mathcal{L}$ is the integrated luminosity of data and $C_i$ is a bin-by-bin correction factor for detector inefficiency, resolution and bin-to-bin migrations.
The term $N^i_{\mathrm{MC,\, det.}}$ is the number of \wzjjew events predicted by the MC at detector level and $N^i_{\mathrm{MC,\, part.}}$ is the corresponding number of MC events predicted at particle level in the fiducial phase space.
Correction factors are calculated using the \MGPy8 MC simulations.
The same control regions, without any \njets or \mjj splitting, and a similar parameterisation of the fit as described in Section~\ref{sec:meth_inclusive} are used.
Correlations between bins of experimental and background systematic uncertainties detailed in Section~\ref{sec:systematics} are taken into account in the simultaneous fit.

\subsection{\wzjj differential measurements}\label{sec:wzjj_diff_method}

Events in the SR are also used to  measure the \wzjj\ differential production cross-section in the VBS fiducial phase space.
The differential detector-level distributions are corrected for detector efficiency and resolution effects 
using an iterative Bayesian unfolding method~\cite{DAgostini:1994zf}, as implemented in the RooUnfold toolkit~\cite{RooUnfold}.
The number of iterations was tuned in order to obtain the best trade-off in minimising both the unfolding bias and the unfolding statistical uncertainty.
Three iterations were used for the unfolding of each variable.
The width of the bins in each distribution is chosen according to the experimental resolution and to the statistical significance of the expected number of events in that bin.
The fraction of signal MC events reconstructed in the same bin as generated is always greater than $40\%$ and around $70\%$ on average.
 
For each  distribution, simulated \wzjj\ events are used to obtain a response matrix that accounts for bin-to-bin migration effects between the reconstruction-level and particle-level distributions.
The \MGPy8 MC samples for \wzjjew, \wzjjint and \wzjjqcd production are added together to model \wzjj production.
Because of differences in event kinematics, the response matrix for separate \wzjjew and \wzjjqcd processes are close but not similar.
To better model the data and to minimise unfolding uncertainties, their predicted cross-sections are rescaled by the signal strengths of $1$ and $0.71$ for the \wzjjew\ and \wzjjnew contributions, respectively, as measured by the maximum-likelihood fit described in Section~\ref{sec:meth_inclusive} applied to data in Section~\ref{sec:results_incl}.
The background contributions from \zzjjqcd, \ttv and \tzj are also rescaled according to the result of the maximum-likelihood fit of Section~\ref{sec:meth_inclusive}, following constraints from data in the CRs.


\section{Systematic uncertainties}
\label{sec:systematics}

Systematic uncertainties in the signal and control regions affecting the shape and normalisation of the BDT score, \mjj and \CRbBDT distributions for the individual backgrounds, as well as the acceptance of the \wzjjew, \wzjjqcd and \wzjjint events and the shape of their templates are considered.
Shape and normalisation uncertainties affecting \wzjjew, \wzjjqcd and \wzjjint contributions in control regions are also considered.
A bootstrapping procedure~\cite{ATL-PHYS-PUB-2021-011} is used to ensure statistically significant variations as a function of considered distributions.

Systematic uncertainties due to the theoretical modelling in the event generator used to evaluate the \wzjjqcd and \wzjjew templates are considered.
Uncertainties due to higher order QCD corrections are evaluated by varying the renormalisation, \muR, and factorisation, \muF, scales independently by factors of two and one-half, removing combinations where the variations differ by a factor of four.
For the \wzjjew process, simulated only at LO in QCD, changes in the definition of \muF are also considered: dedicated MC studies show that depending on the definition of \muF the shape of the BDT score distribution is subject to changes not covered by the scaling of \muF by factors half or one-half.
In the SR, these uncertainties are of up to $\pm 5\%$ and of $\pm 20\%$ on the shape of the BDT score distribution for \wzjjqcd and \wzjjew events, respectively.
The uncertainties due to the PDF and the $\alphas$ value used in the PDF determination are evaluated using the \texttt{PDF4LHC} prescription~\cite{Butterworth:2015oua} and are of the order of $1\%$ to $2\%$ in shape.
A global modelling uncertainty in the \wzjjqcd background template that includes effects of the parton shower model is estimated by comparing predictions of the BDT score distribution in the signal region from the \MGPy8 and \sherpa MC event generators.
The difference between the predicted shapes of the BDT score distribution from the two generators is considered as an uncertainty and affects the shape of the BDT score distribution of \wzjjqcd events by up to $30\%$ at larger values of the BDT score.
A parton-shower uncertainty in the \wzjjew signal template is estimated by comparing predictions of the BDT score distribution in the signal region from MC simulations from \madgraph interfaced to \pythia or \herwig.
This parton-shower uncertainty affects the shape of the BDT score distribution of \wzjjew events by at most $10\%$ at lower values of the BDT score.
For the \wzjjint template, uncertainties due to variations of the renormalisation and factorisation scales and due to the PDF and to the $\alphas$ value used in the PDF determination are considered.
They affect the shape of the template by approximately $3\%$.

For the measurement of integrated and differential \sigew and \sigqcd cross-sections using a combined maximum-likelihood fit, all the uncertainties related to higher order QCD corrections, parton shower and modelling are considered as uncorrelated between bins in \njets or \mjj in the SR and between the SR and the control regions.
This leads to the most conservative fit approach for these uncertainties.

For the measurements of  \sigew, \sigqcd and \sigtot differential cross-sections, uncertainties in the unfolding due to imperfect modelling of the data by the MC simulation are evaluated using a data-driven method~\cite{Malaescu:2011yg}, where the MC differential distribution is corrected to match the data distribution and the resulting weighted MC distribution at reconstruction level is unfolded with the response matrix used in the data unfolding.
The new unfolded distribution is compared with the weighted MC distribution at generator level and the difference is taken as the systematic uncertainty.
For \sigew and \sigqcd cross-sections, the uncertainties obtained range from $4\%$ to $9\%$.
For \sigtot cross-sections, the uncertainties range from $0.1\%$ to $16\%$ depending on the resolution of the unfolded observables and on the quality of its description by \MGPy8.
They cover residual modelling uncertainties in the unfolded observables, which could still be present after the adjustment to data of the  normalisation of \wzjjqcd and \wzjjew contributions.
It is also verified that, by construction, these uncertainties cover for possible EFT contributions being present in the data.
An additional uncertainty is derived to account for more subtle differences between the \MGPy8 and \sherpa generators (e.g.\ hadronisation models, additional soft objects, mis-modelling  in other kinematic variables).
The \sherpa generator is used to unfold the data and deviation from the nominal result is taken as the uncertainty.
To remove effects already accounted for in the data-driven method, the \sherpa distributions were first reweighed to match \MGPy8 distributions at particle level.

Systematic uncertainties affecting the reconstruction and energy calibration of jets, electrons and muons are propagated through the analysis.
The dominant sources of uncertainties are the jet energy scale calibration, including the modelling of pile-up.
The uncertainties in the jet energy scale are obtained from $\sqrt{s} = 13$~\TeV\ simulations and {\textit{in situ}} measurements~\cite{JETM-2018-05}.
The uncertainties in the jet energy resolution~\cite{JETM-2018-05}, in the suppression of jets originating from pile-up~\cite{Aad:2015ina}, in the $b$-tagging efficiency and in the mistag rate~\cite{FTAG-2019-07} are also considered.
The uncertainty in the \MET measurement is estimated by propagating the uncertainties in the transverse momenta of hard physics objects and by applying momentum scale and resolution uncertainties to the track-based soft term~\cite{ATLAS-CONF-2018-023}. 
A variation in the pile-up reweighting of MC events is included in order to cover the uncertainty in the ratio of the predicted and measured $pp$ inelastic cross-sections~\cite{Aaboud:2016mmw}.

The uncertainties due to lepton reconstruction, identification, isolation requirements and trigger efficiencies as well as in the lepton momentum scale and resolution are assessed using tag-and-probe methods in $Z \to \ell \ell$ events~\cite{ATLAS:2019qmc,ATLAS:2016lqx,ATLAS:2020auj}.

The uncertainty in the amount of background from misidentified leptons takes into account the limited number of events in the control regions as well as the difference in background composition between the control region used to determine the lepton misidentification rate and the control regions used to estimate the yield in the signal region.
This results in an uncertainty of about $25\%$ in the total  misidentified-lepton background yield and in the shape of the differential distributions of the reducible background events.

The uncertainty due to irreducible background sources that are not constraint by data in the dedicated CRs is evaluated by propagating the uncertainty in their MC cross-sections. These are $20\%$ for $VVV$~\cite{ATL-PHYS-PUB-2016-002} 
and $25\%$ for \zzjjew to account for the potentially large impact of scale variations.

The uncertainty in the combined $2015$--$2018$ integrated luminosity is $0.83\%$~\cite{DAPR-2021-01}, obtained using the LUCID-2 detector~\cite{LUCID2} for the primary luminosity measurements, complemented by measurements using the inner detector and calorimeters.

The effect of the systematic uncertainties on the final results after the maximum-likelihood fit is shown in Table~\ref{table:sysOnMu_summary} where the breakdown of the contributions to the uncertainties in the measured fiducial cross-section \sigew and \sigqcd is presented.
The individual sources of systematic uncertainty  are combined into theory modelling and experimental categories.
As shown in the table, the systematic uncertainties in  the modelling of the \wzjjew and \wzjjqcd processes play a dominant role, followed by the uncertainties in the jet reconstruction and calibration.
Systematic uncertainties in the missing transverse momentum computation arise directly from the momentum and energy calibration of jets, electrons and muons and are included in the respective lines of Table~\ref{table:sysOnMu_summary}.
Systematic uncertainties in the modelling of the other backgrounds are also considered.

\begin{table}[t]
\caption{Grouped impact of systematic uncertainties in the measured integrated fiducial cross-sections \sigew and \sigqcd. The uncertainties are reported as percentages. The total systematic uncertainty does not correspond directly to the sum in quadrature of the distinct sources due to correlations introduced by the fit, in particular between \sigew and \sigqcd.
\label{table:sysOnMu_summary}}
\centering

\begin{tabular}{l r@{\,}l r@{\,}l }
\toprule
Source & \multicolumn{2}{l}{$\frac{\Delta\sigew }{ \sigew}$ [\%]} & \multicolumn{2}{l}{$\frac{\Delta\sigqcd }{ \sigqcd}$ [\%]}  \\
\midrule
\wzjjew  theory modelling & 7 &  & 1 & .8 \\
\wzjjqcd theory modelling & 2 & .8 & 8 &  \\
\wzjjew and \wzjjqcd interference & 0 & .35 & 0 & .6 \\
PDFs & 1 & .0 & 0 & .06 \\
\midrule
Jets & 2 & .3 & 5 &  \\
Pile-up & 1 & .1 & 0 & .6 \\
Electrons & 0 & .8 & 0 & .8 \\
Muons & 0 & .9 & 0 & .9 \\
$b$-tagging & 0 & .10 & 0 & .11 \\
MC statistics & 1 & .9 & 1 & .2 \\
Misid. lepton background & 2 & .3 & 2 & .3 \\
Other backgrounds & 0 & .9 & 0 & .23 \\
\midrule
Luminosity & 0 & .7 & 0 & .9 \\
All systematics & 16 &  & 12 &  \\
Statistics & 10 &  & 6 &  \\
\midrule
Total & 19 &  & 13 &  \\
\toprule
\end{tabular}


\end{table}


\section{Results}
\label{sec:xsection_results}
 
\subsection{\wzjjew and \wzjjnew integrated measurements}\label{sec:results_incl}


\begin{figure}[t]
\centering
\subfloat[]{
\includegraphics[width=0.4\textwidth]{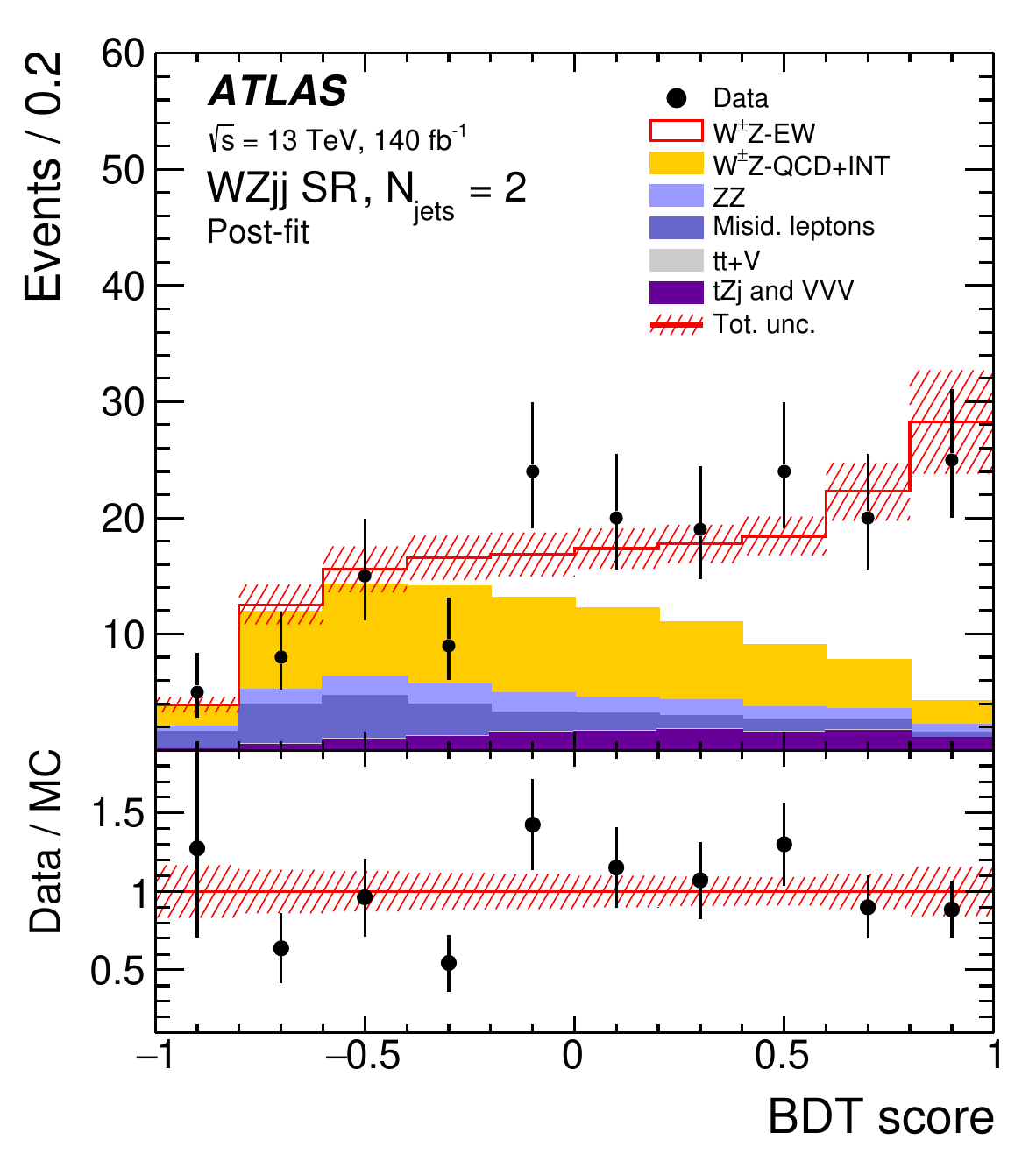}
}
\subfloat[]{
\includegraphics[width=0.4\textwidth]{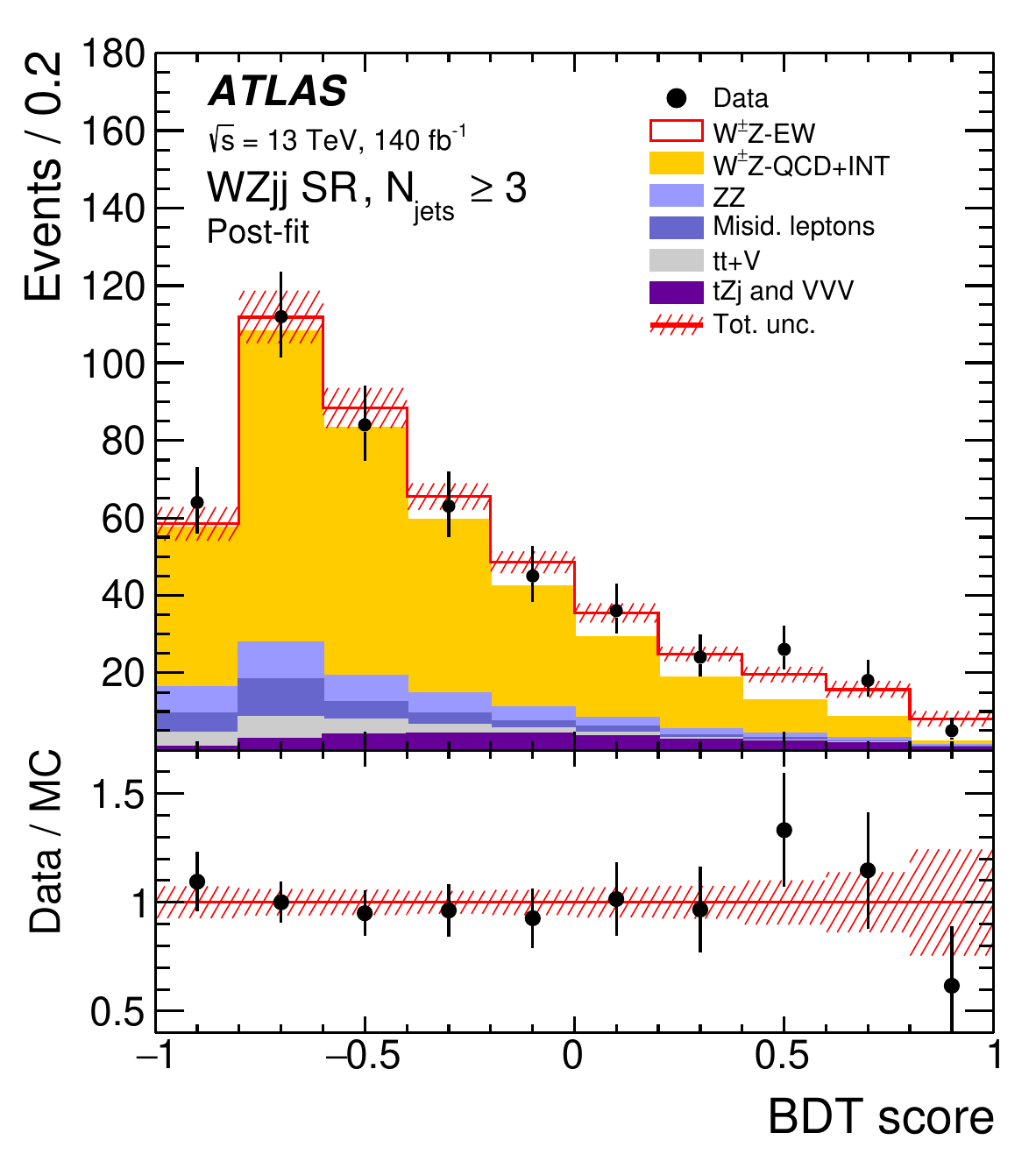}
}\\
\subfloat[]{
\includegraphics[width=0.4\textwidth]{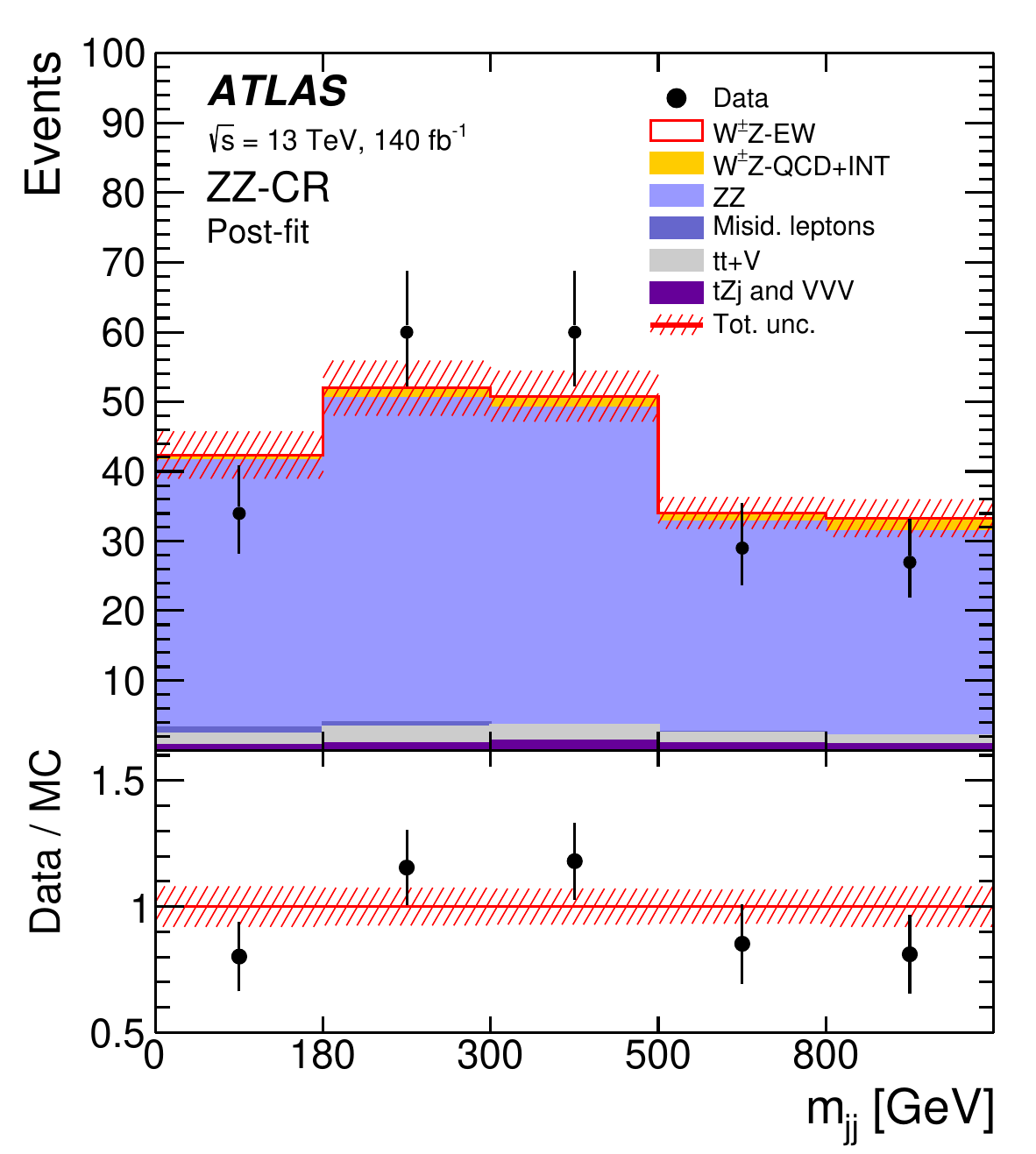}
}
\subfloat[]{
\includegraphics[width=0.4\textwidth]{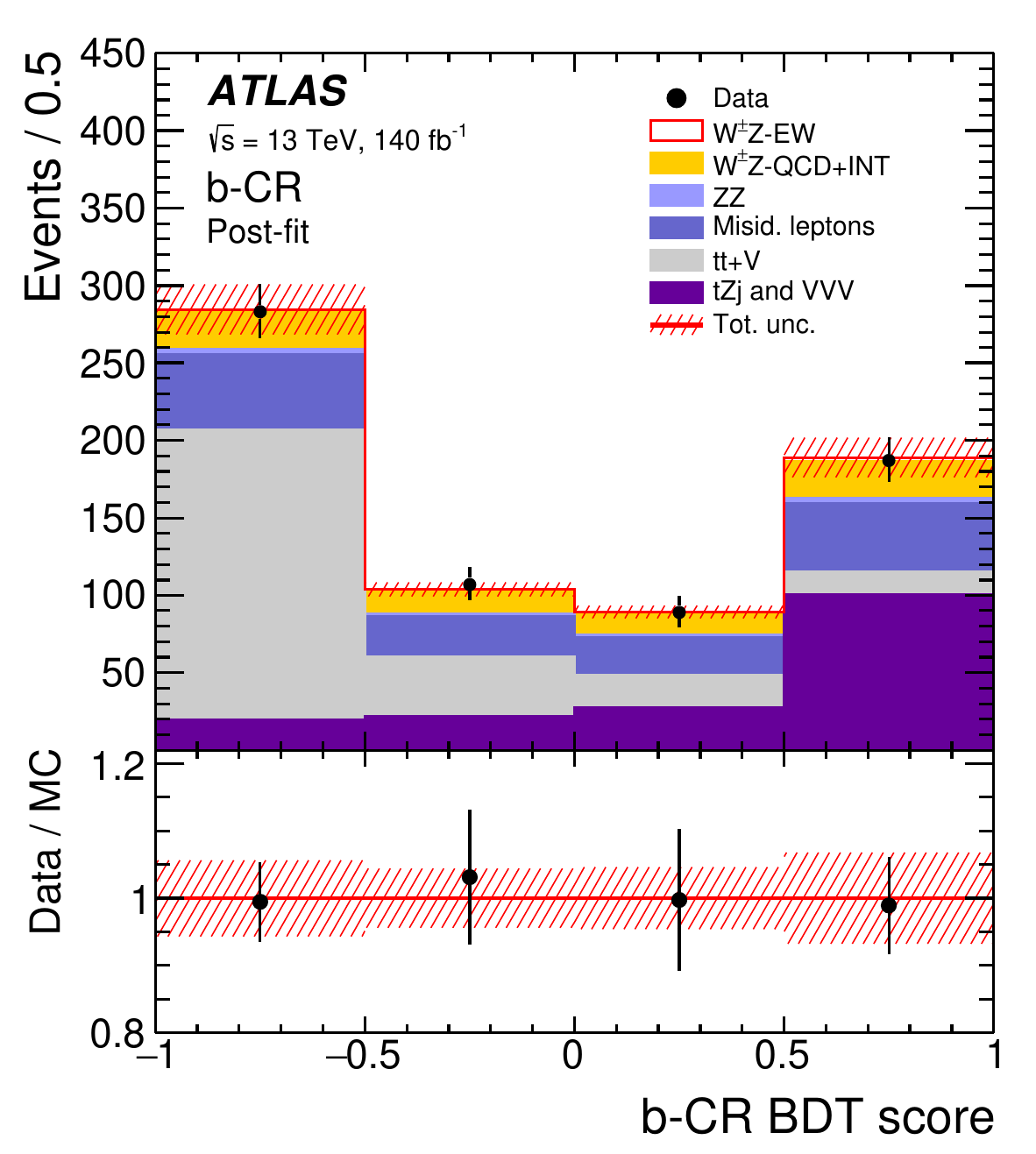}
}
 
\caption{Post-fit distributions of (a) the BDT score distribution in the signal region, for $\njets = 2$ and (b) $\njets \ge 3$ categories, (c) \mjj in the \ZZCR control region, (d) the \CRbBDT score in the \bCR.
Signal and backgrounds are normalised to the expected number of events after the fit.
The uncertainty band around the MC expectation includes all systematic uncertainties as obtained from the fit.
}
\label{fig:BDTScore}
\end{figure}

\begin{table}[t] 
\caption{Observed and expected numbers of events in the \wzjj signal region and in the two control regions, after the fit.
The expected number of \wzjjew, \wzjjqcd and \wzjjint events and the estimated number of background events from the other processes are shown.
The sum of the backgrounds containing misidentified leptons is labelled `Misid. leptons'.
The total correlated post-fit systematic uncertainties are quoted.
\label{tab:VBSyield_postfit}}
\centering

\begin{tabular}{l r@{\,}l!{$\pm$}r@{\,}l r@{\,}l!{$\pm$}r@{\,}l r@{\,}l!{$\pm$}r@{\,}l r@{\,}l!{$\pm$}r@{\,}l }
\toprule
& \multicolumn{4}{c}{SR, $\Njets = 2$}  & \multicolumn{4}{c}{SR, $\Njets \ge 3$}  & \multicolumn{4}{c}{$b$-CR} & \multicolumn{4}{c}{$ZZ$-CR} \\
\midrule
Data & \multicolumn{4}{c}{169}  & \multicolumn{4}{c}{477}  &   \multicolumn{4}{c}{666} & \multicolumn{4}{c}{210}  \\
Total pred. & 170& & 13&  & 476& & 22&  &   667& & 26& & 212& & 14&  \\
\midrule
$WZjj$-EW & 68& & 14&  & 55& & 18&  &   4&.84 & 0&.27 & 0&.724 & 0&.014  \\
$WZjj$-QCD & 58& & 16&  & 307& & 27&  &   77& & 18& & 6&.3 & 0&.7  \\
$WZjj$-INT & 0&.9 & 0&.4  & 4&.4 & 2&.3  &   0&.57 & 0&.29 & 0&.22 & 0&.11  \\
$t\bar{t}+V$ & 0&.59 & 0&.10  & 18&.3 & 2&.4  &   262& & 34& & 9&.0 & 1&.3  \\
$tZj$ & 11&.0 & 1&.9  & 25& & 5&  &   169& & 30& & 0&.54 & 0&.09  \\
$ZZ$-QCD & 10&.3 & 1&.0  & 34&.6 & 3&.2  &   10&.1 & 0&.5 & 171& & 15&  \\
$ZZ$-EW & 1&.9 & 0&.4  & 3&.7 & 0&.9  &   0&.21 & 0&.05 & 19& & 5&  \\
$VVV$ & 0&.41 & 0&.10  & 2&.0 & 0&.5  &   0&.39 & 0&.10 & 4&.2 & 1&.0  \\
Misid. leptons & 18& & 4&  & 27& & 6&  &   143& & 35& & 1&.7 & 0&.5  \\
\bottomrule
\end{tabular}


\end{table}

The post-fit distributions of the BDT score in the signal region for events with $\Njets = 2$ and $\Njets \ge 3$ are presented in Figures~\ref{fig:BDTScore}(a) and (b), respectively.
The background normalisations, signal normalisation and nuisance parameters are adjusted by the profile-likelihood fit.
The post-fit distributions of \mjj and of the \CRbBDT score in the \ZZCR and \bCR, respectively, are also presented in Figures \ref{fig:BDTScore}(c) and (d).
The corresponding post-fit event yields are detailed in Table~\ref{tab:VBSyield_postfit}.
The normalisation parameters of the \ttV and $ZZ$ backgrounds constrained by data in the control and signal regions are measured to be  $\muttV = 0.91 \pm 0.13$, $\mutZ = 1.26 \pm 0.22$ and $\muZZ = 1.07 \pm 0.11$. 
No pull of nuisance parameters by more than $0.5\sigma$ is observed.

The integrated fiducial cross-sections \sigew and \sigqcd, for a given decay channel $\wz \rightarrow \ell^{'\pm} \nu \ell^+ \ell^-$, where $\ell$ and $\ell^{'}$ are either an electron or a muon, are measured to be

\[
\begin{array}{rcl}
\sigew &=& 0.368 \; \pm 0.037 \,\mathrm{(stat.)} \; \pm 0.059 \,\mathrm{(syst.)} \; \pm 0.003 \,\mathrm{(lumi.)} \; \mathrm{fb} \, \\
&=& 0.37 \; \pm 0.07 \; \mathrm{fb} ,\\
\sigqcd &=& 1.093 \; \pm 0.066 \,\mathrm{(stat.)} \; \pm 0.131 \,\mathrm{(syst.)} \; \pm 0.009 \,\mathrm{(lumi.)} \; \mathrm{fb} \, \\
&=& 1.09 \; \pm 0.14 \; \mathrm{fb} ,\\
\end{array}
\]
 
where the uncertainties correspond to statistical, systematic, and luminosity uncertainties, respectively.
Table~\ref{table:sysOnMu_summary} shows the main sources of uncertainty in the measured cross-sections.
The precision of the measurements is limited by the systematic uncertainties, and in particular by the model uncertainties of the MC event generators.
The total systematic uncertainties are also increased by existing correlations between the \sigew and \sigqcd measured cross-sections, which are of the order of $36\%$.
The corresponding predictions from \MGPy8 are
 
\[
\begin{array}{rcl}
\sigew^{\MGPy8} &=& 0.370 \; \pm 0.001 \,\mathrm{(stat.)} \; \pm 0.006 \,\mathrm{(PDF)} \; ^{+ 0.030}_{- 0.026} \,\mathrm{(scale)} \; \mathrm{fb} \, ,\\
\sigqcd^{\MGPy8} &=& 1.537 \; \pm 0.009 \,\mathrm{(stat.)} \; \pm 0.016 \,\mathrm{(PDF)} \; ^{+ 0.087}_{- 0.149} \,\mathrm{(scale)} \; \mathrm{fb} \, ,\\
\end{array}
\]
 
where the uncertainties correspond to statistical, PDF and QCD scale uncertainties, respectively, calculated as detailed in Section~\ref{sec:systematics}.
The measured values of \sigew and \sigqcd are also compared to the predictions from \MGPy8 and \SHERPA[2.2.12] in Figure~\ref{fig:2DContour_sigew_sigqcd_All_data}.
A good agreement of the MC predictions with the measured \wzjjew integrated cross-section is observed, while the \wzjjqcd integrated cross-section is measured to be lower than the prediction from both MC event generators by a factor of $0.7$.
Considering the theory uncertainty in the \MGPy8 prediction, the present measurement and the \MGPy8 prediction agree within $1.8\sigma$.
A similar observation regarding the modelling of  \wzjjqcd production was also made by the ATLAS Collaboration using a different measurement phase space~\cite{ATLAS:2023sua}.
It is also consistent with measurements done in previous ATLAS publications dedicated to \wzjj and \wz final states~\cite{WZVBS36fb,WZPol36fb}.

\begin{figure}[t]
\centering
\includegraphics[width=0.49\textwidth]{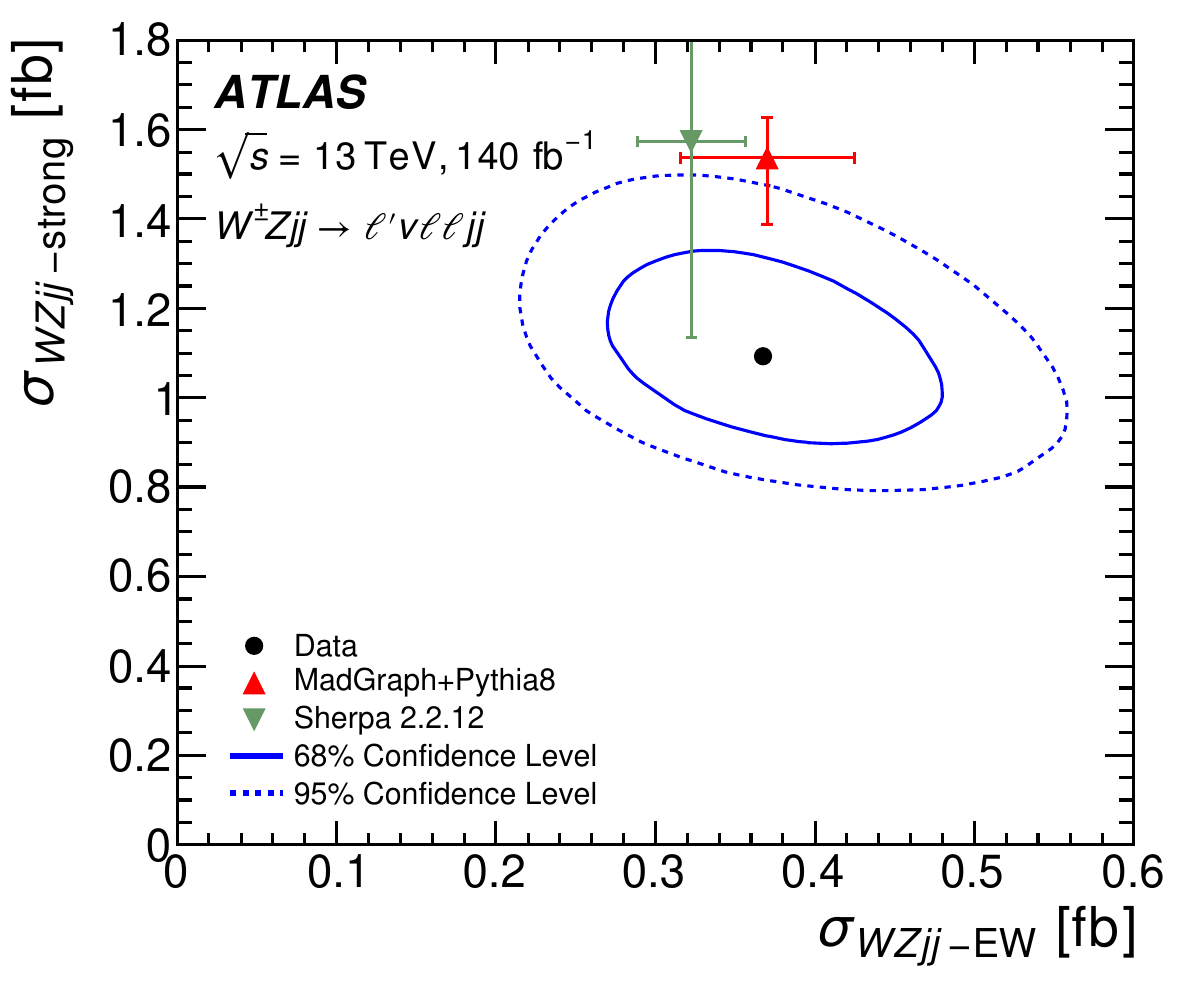}
\caption{The measured \sigew and \sigqcd integrated cross-sections compared with predictions from \MGPy8 (upward pointing triangle) and \sherpa 2.2.12 (downward pointing triangle).
The full and dashed contours around the data points correspond to $68$\% and $95$\% CL, respectively.
}
\label{fig:2DContour_sigew_sigqcd_All_data}
\end{figure}

The \wzjj integrated cross-section in the fiducial phase space is also measured from the fit to be $\sigtot = 1.462 \;\pm 0.133 \, \mathrm{fb} \; =  \; 1.462 \pm  0.063 \, \mathrm{(stat.)} \;\pm 0.118 \, \mathrm{(sys.)} \;\pm 0.012 \, \mathrm{(lumi.)} \, \mathrm{fb}$.
The corresponding prediction from \MGPy8 for strong and electroweak production and including interference effects is
$1.907 \pm 0.009 \,\mathrm{(stat.)} \pm 0.022 \,\mathrm{(PDF)} ^{+0.117}_{-0.175} \,\mathrm{(scale)}  \; \mathrm{fb}$.


\subsection{\wzjjew and \wzjjnew differential measurements}


Dividing the SR in events with $\Njets = 2$ and $\Njets \ge 3$ the \sigew and \sigqcd production cross-sections of these two categories of events are measured.
The measurements are compared in Figure~\ref{fig:2DContour_sigew_sigqcd_Nj_data} to the prediction from \MGPy8 and \sherpa~2.2.12.
For $\Njets \ge 3$, the predicted \sigew cross-sections are in good agreement with the measured value while the predicted \sigqcd cross-section lie within about 2$\sigma$ of the measurement.
However, for $\Njets = 2$ the measured \sigqcd cross-section is lower by a factor of two than the value predicted by both \MGPy8 and \sherpa.
The predicted values of \sigew are found to be in agreement within 1$\sigma$ of the measured value.
The ratio $R_{2/3}$ of the number of events with  $\Njets = 2$ to the number of events with $\Njets \ge 3$ is also extracted from data by the simultaneous fit of the two categories to be $\rew = 1.70 \pm 0.71$ and $\rqcd = 0.21 \pm 0.06$ for \wzjjew and \wzjjqcd events, respectively.
In comparison, the values predicted by \MGPy8 (\sherpa) are $\rew = 1.43 ^{+0.06}_{-0.02} \, (1.67 \pm 0.13)$ and $\rqcd = 0.36 ^{+0.02}_{-0.04} \, (0.38 \pm 0.03)$, respectively.

\begin{figure}[t]
\centering
\subfloat[]{\includegraphics[width=0.49\textwidth]{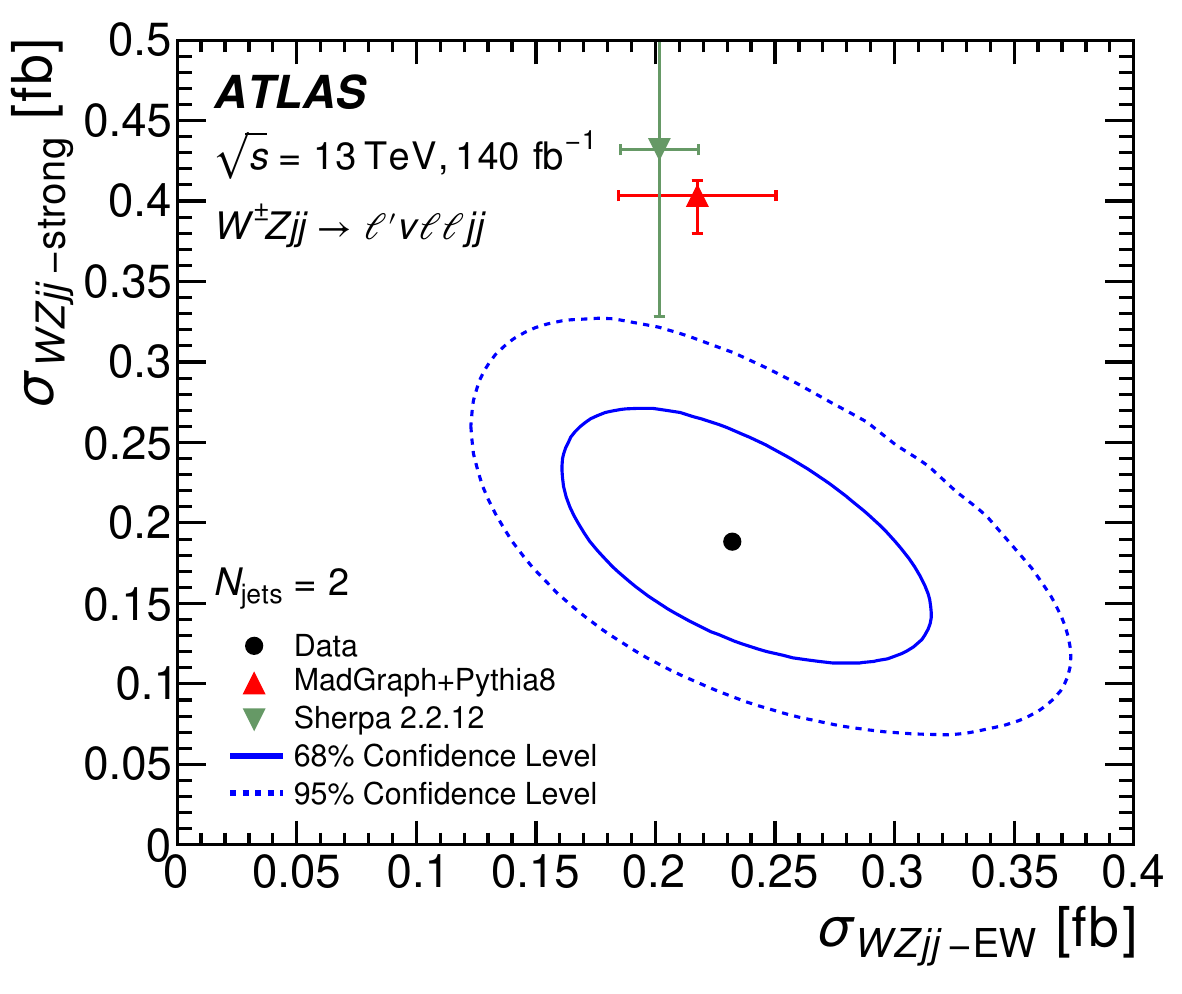}}
\subfloat[]{\includegraphics[width=0.49\textwidth]{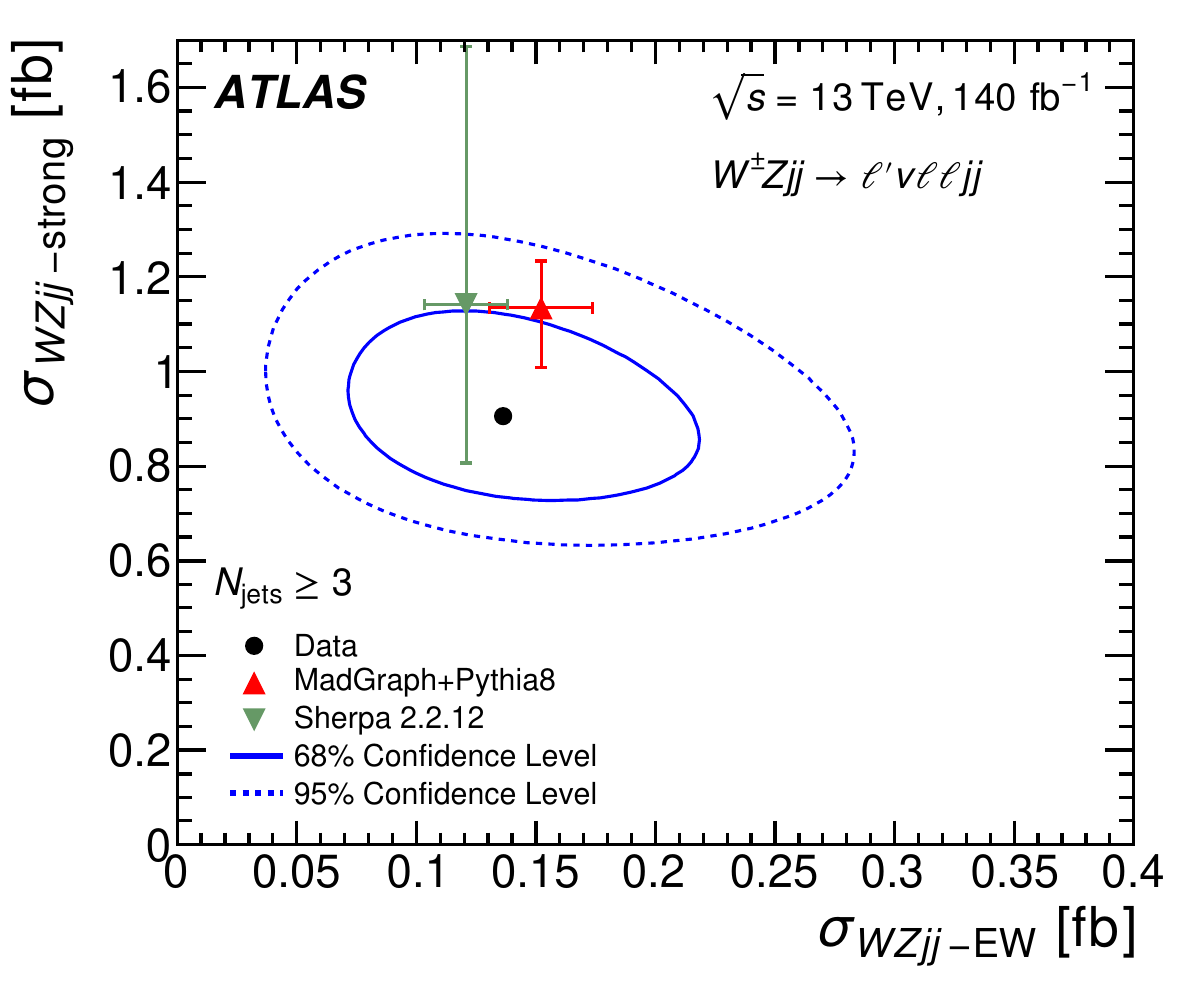}}
\caption{The measured \sigew and \sigqcd cross-sections (a) for $\Njets = 2$  and (b) $\Njets \ge 3$  compared with predictions from \MGPy8 (upward pointing triangle) and \sherpa~2.2.12 (downward pointing triangle).
The full and dashed contours around the data points correspond to $68$\% and $95$\% CL, respectively.}
\label{fig:2DContour_sigew_sigqcd_Nj_data}
\end{figure}

The \sigew and \sigqcd production cross-sections are measured differentially in three bins of \mjj.
The measurements are compared in Figure~\ref{fig:2DContour_sigew_sigqcd_mjj_data} to the prediction from \MGPy8 and \sherpa~2.2.12.
For $500 < \mjj < 1300$~\GeV, the MC predictions are found to overestimate the measured \sigqcd value by a factor $1.4$.
For \mjj values above $1300$~\GeV, the MC predictions agree within approximately $2\sigma$ with the measurements, for both \sigew and \sigqcd.

\begin{figure}[t]
\centering
\subfloat[]{\includegraphics[width=0.45\textwidth]{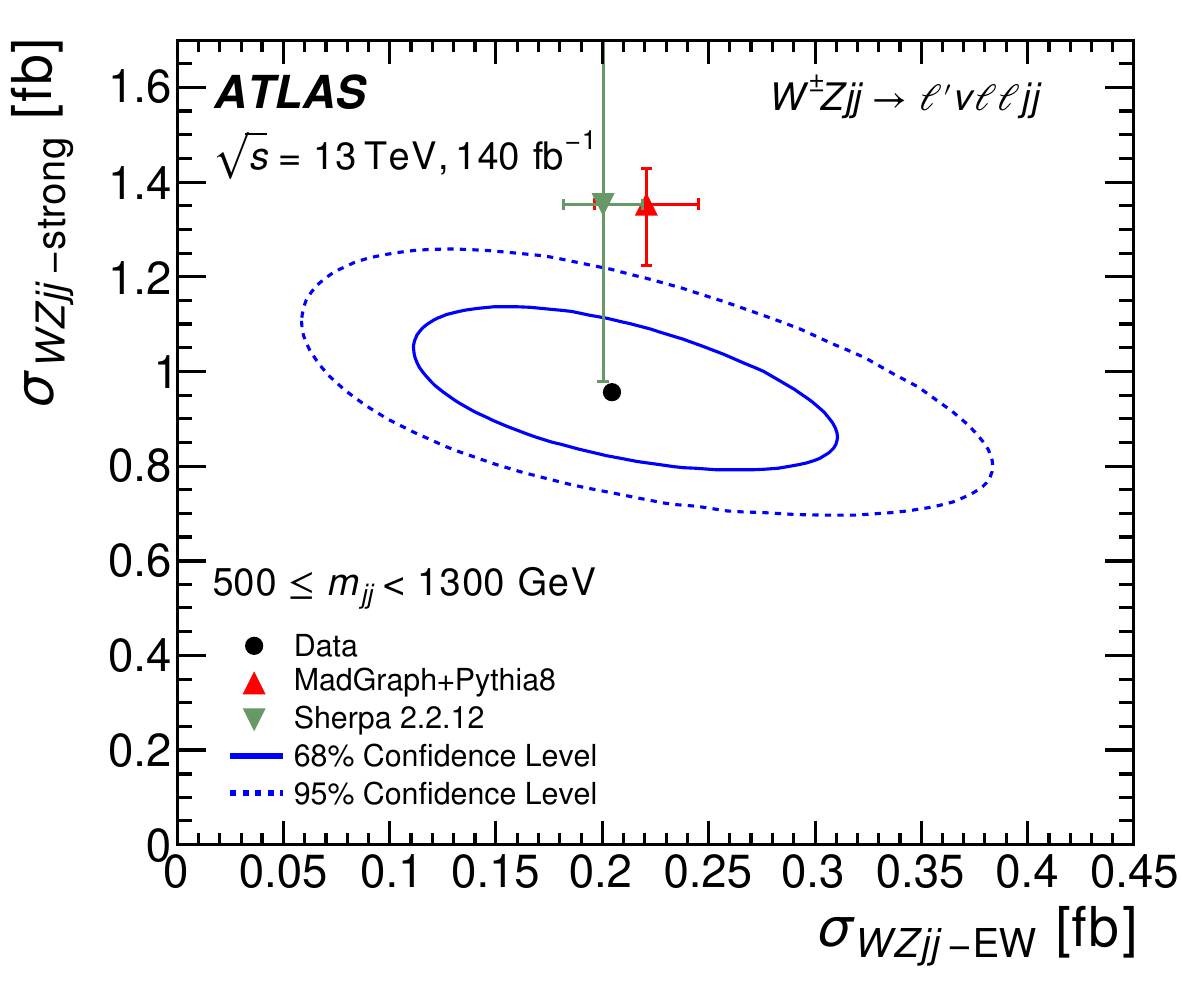}}
\subfloat[]{\includegraphics[width=0.45\textwidth]{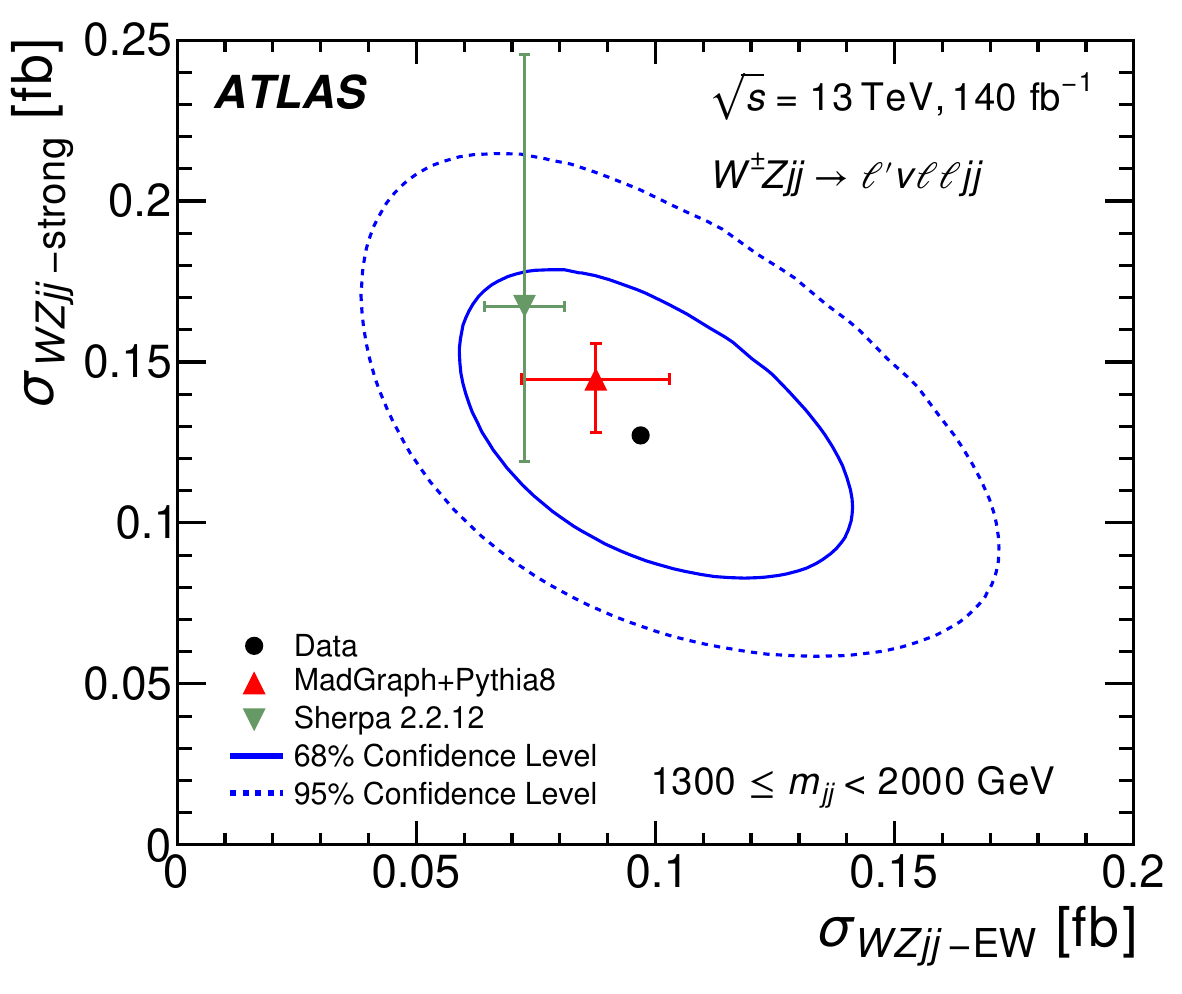}}\\
\subfloat[]{\includegraphics[width=0.45\textwidth]{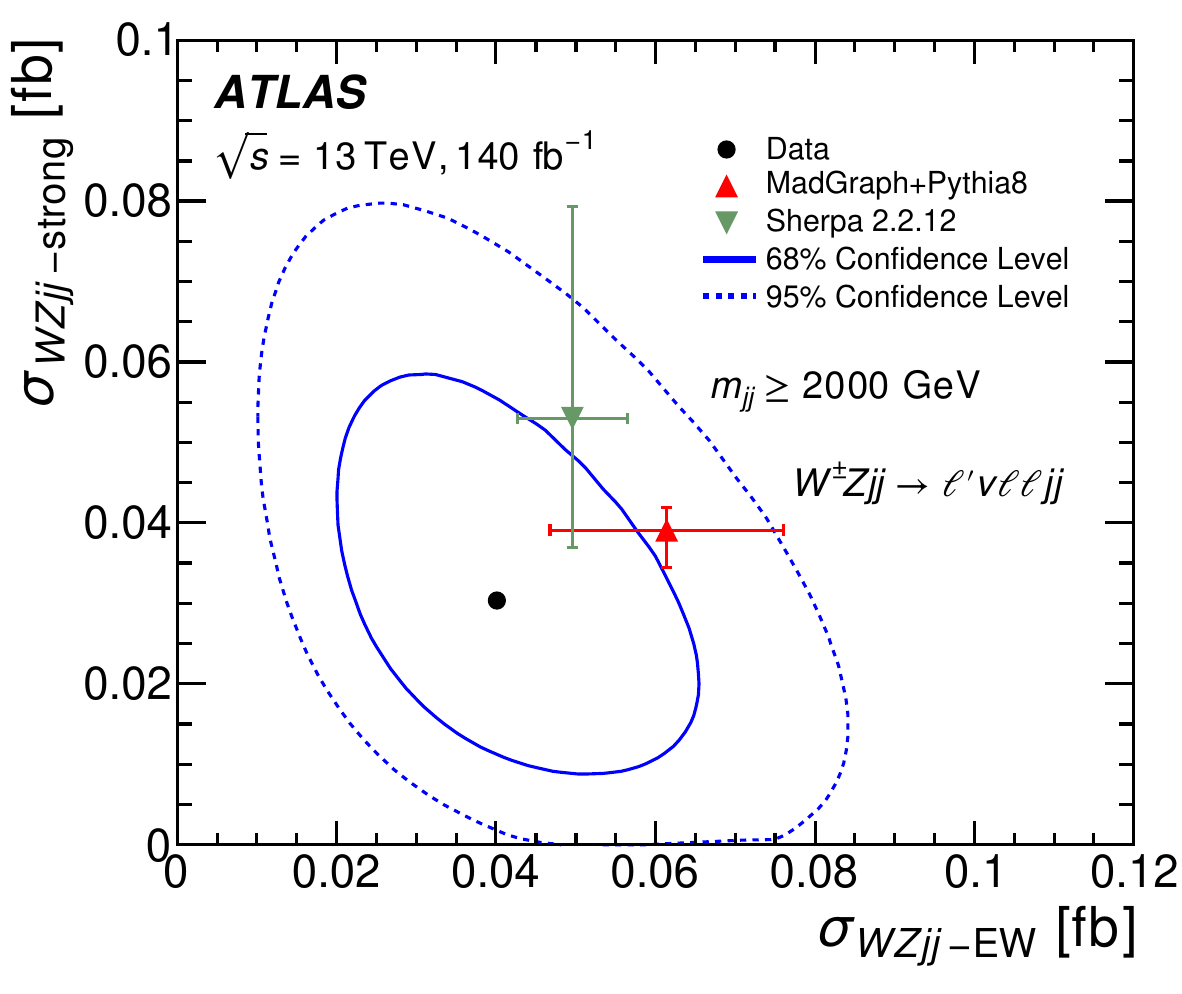}}
\caption{The measured \sigew and \sigqcd cross-sections per \mjj bin, (a) $500 \leq \mjj < 1300$~\GeV, (b) $1300 \leq \mjj < 2000$~\GeV, (c) $\mjj \geq 2000$~\GeV, compared with predictions from \MGPy8 (upward pointing triangle) and \sherpa~2.2.12 (downward pointing triangle).
The full and dashed contours around the data points correspond to $68$\% and $95$\% CL, respectively.}
\label{fig:2DContour_sigew_sigqcd_mjj_data}
\end{figure}


\subsection{\wzjj differential measurements}\label{sec:wzjj_diff_results}


Differential measurements of \wzjj production cross-section are performed as a function of three variables sensitive to anomalies in the quartic gauge coupling in \wzjj events~\cite{ATLASWZ_8TeV}, namely the scalar sum of the transverse momenta of the three charged leptons associated with the $W$ and $Z$ bosons \sptl, the difference in azimuthal angle \dPhiWZ\ between the $W$ and $Z$ bosons' directions, and the transverse mass of the \wz system \mtwz, defined following Ref.~\cite{ATLASWZ_8TeV}.
These are presented in Figure~\ref{fig:diffXSection:SptL_dPhiWZ}.
 
Measurements are also performed as a function of variables related to the kinematics of jets.
The exclusive multiplicity of jets with  $p_\textrm{T} > 40$~\GeV, \njetsf, the absolute difference in rapidity between the two tagging jets \dYjj, the invariant mass of the tagging jets \mjj, the exclusive multiplicity \NjGap of jets with \mbox{$p_\textrm{T} > 25$~\GeV}\ in the gap in $\eta$ between the two tagging jets, and the azimuthal angle between the two tagging jets \dPhijj are shown in Figure~\ref{fig:diffXSection:dYjj_mjj}.
 
Measurements related to the jet activity in the gap between the two tagging jets are also performed using the subset of events in the SR where a third jet with $\pt > 25$~\GeV is present.
The exclusive multiplicity \NjGap of jets with \mbox{$p_\textrm{T} > 25$~\GeV}\ in the gap in $\eta$ between the two tagging jets and the Zeppenfeld variable~\cite{Rainwater:1996ud}, \zj3, for the third jet with \mbox{$p_\textrm{T} > 25$~\GeV}\ are presented in Figure~\ref{fig:diffXSection:3rdjet}.
The \zj3 variable is defined as
 
\begin{equation}
\zj3 = \left | \frac{y_{j_3} - \frac{1}{2} \left( y_{j_1} + y_{j_2}  \right)}{y_{j_1} - y_{j_2}} \right | \; ,
\end{equation}
 
where $y_{j_i}$ is the rapidity of the jet $i$.

Finally, a differential measurement of \wzjj production cross-section is performed as a function of the \BDTscore and is presented in Figure~\ref{fig:diffXSection:BDT}.
The particle-level \BDTscore prediction is evaluated using the same BDT as trained at detector level and evaluated using particle-level input observables.

Total uncertainties in the measurements are dominated by statistical uncertainties.
The differential measurements are compared with the prediction from \MGPy8, after having rescaled the separate \wzjjqcd and \wzjjew components respectively to the inclusive \sigqcd and \sigew cross-sections obtained from the profile-likelihood fit to data.
Interference effects between the \wzjjqcd and \wzjjew processes are incorporated in the \wzjjqcd prediction.
The measurements are also compared with the predictions from \sherpa~2.2.12.

\begin{figure}[t]
\centering
\subfloat[]{\includegraphics[width=0.45\textwidth]{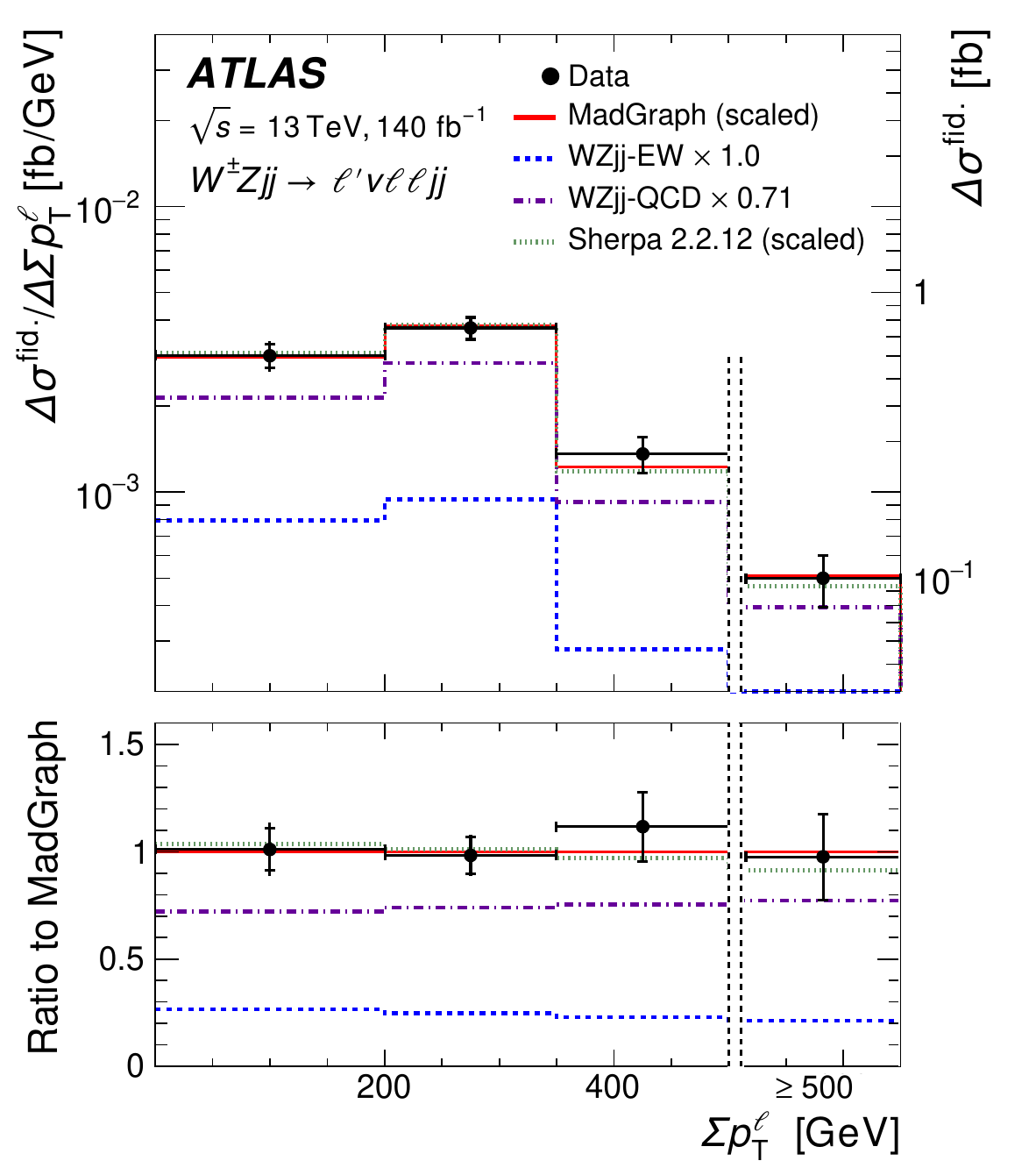}}
\subfloat[]{\includegraphics[width=0.45\textwidth]{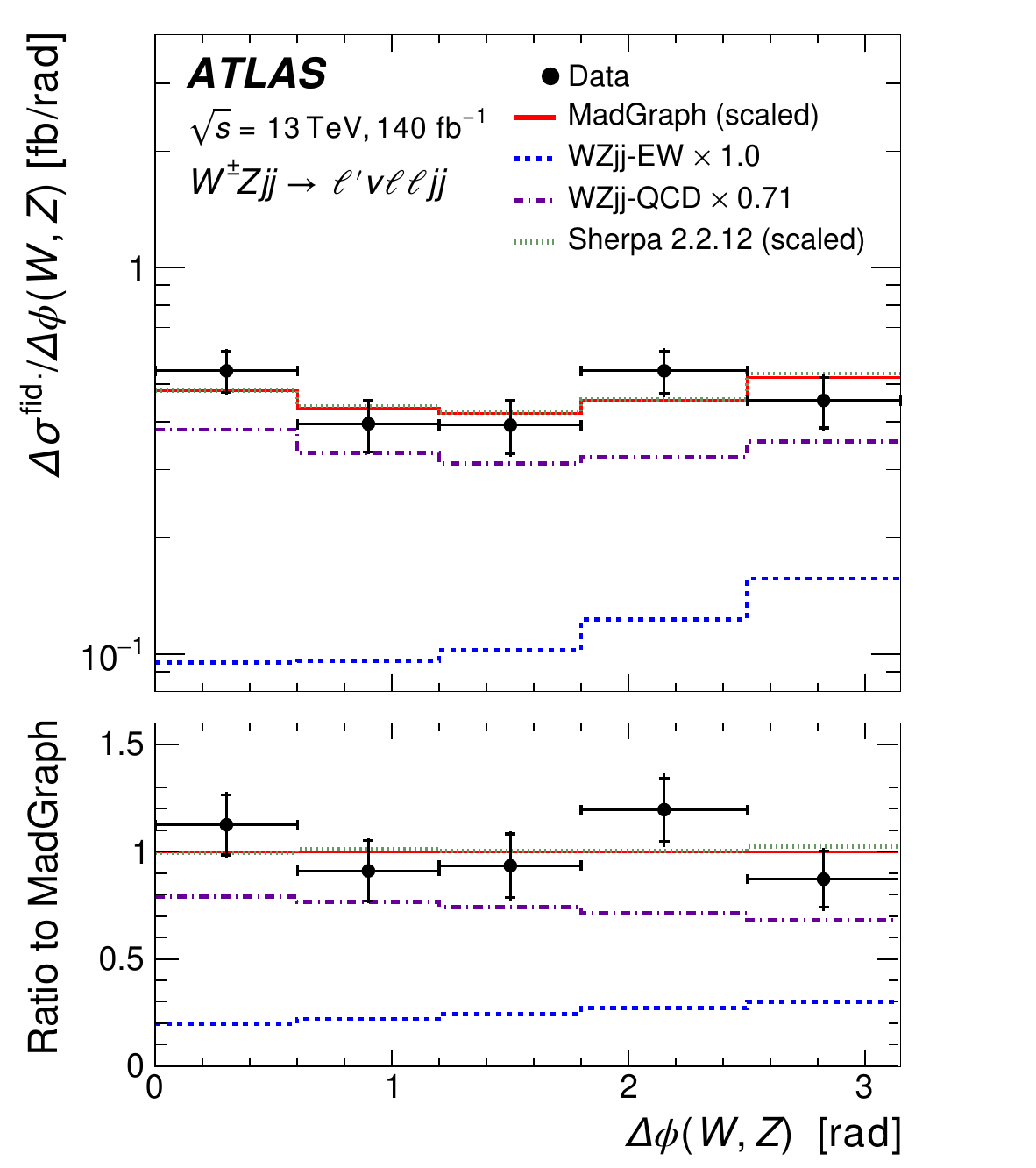}}\\
\subfloat[]{\includegraphics[width=0.45\textwidth]{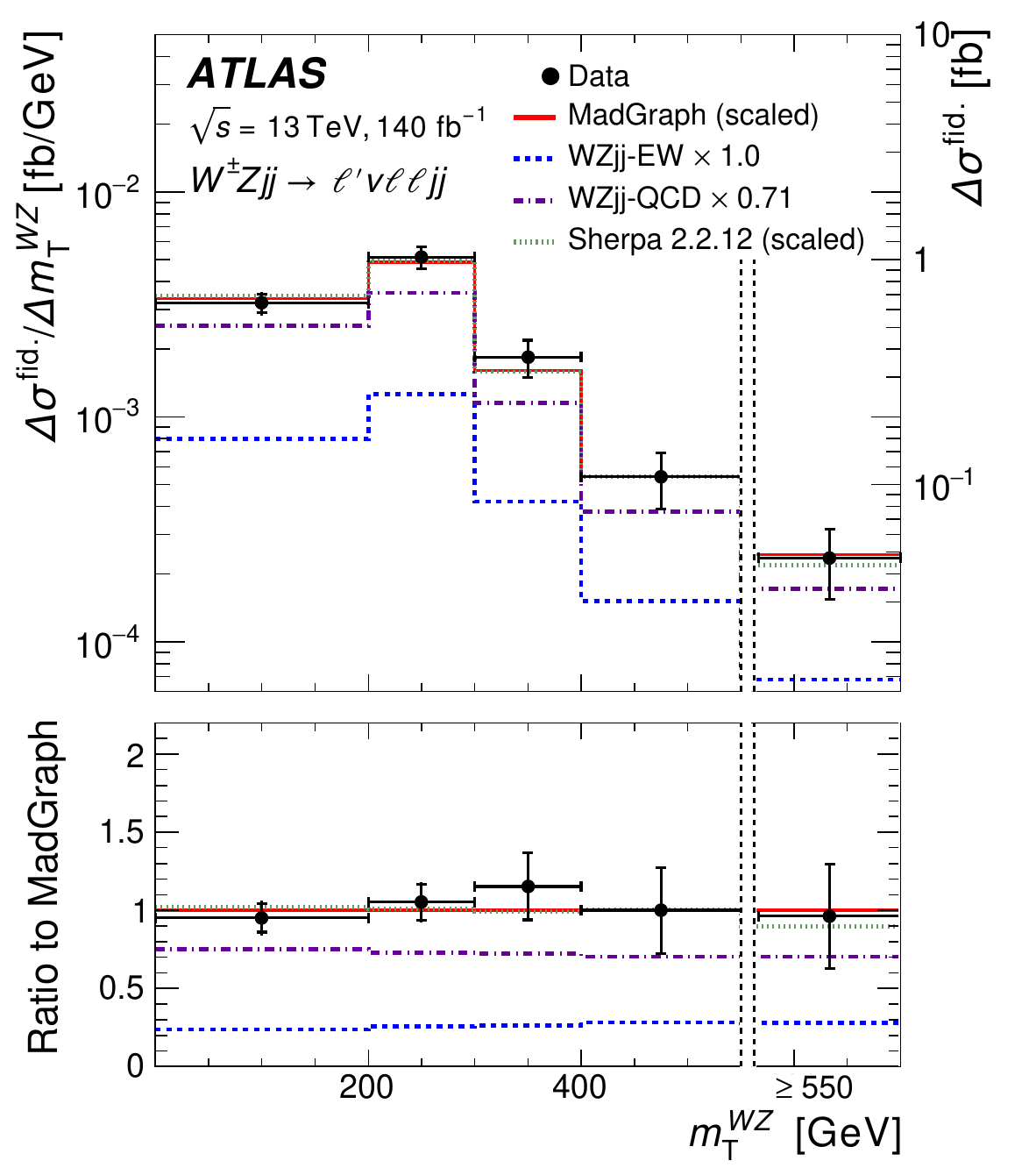}}
 
\caption{The measured \wzjj differential cross-section in the VBS fiducial phase space as a function of (a) \sptl, (b) \dPhiWZ and (c) \mtwz.
The inner and outer error bars on the data points represent the statistical and total uncertainties, respectively.
The measurements are compared with the sum of the rescaled \wzjjqcd and \wzjjew predictions from \MGPy8  (solid line) and \sherpa 2.2.12 (dotted line).
The \wzjjew and \wzjjqcd contributions are also represented by dashed and dashed-dotted lines, respectively.
In (a) and (c), the right $y$-axis refers to the last cross-section point, separated from the others by a vertical dashed line, as this last bin is integrated up to the maximum value reached in the phase space.
The lower panels show the ratios of the data to the predictions from \MGPy8.
}
\label{fig:diffXSection:SptL_dPhiWZ}
\end{figure}

\begin{figure}[t]
\centering
\subfloat[]{\includegraphics[width=0.45\textwidth]{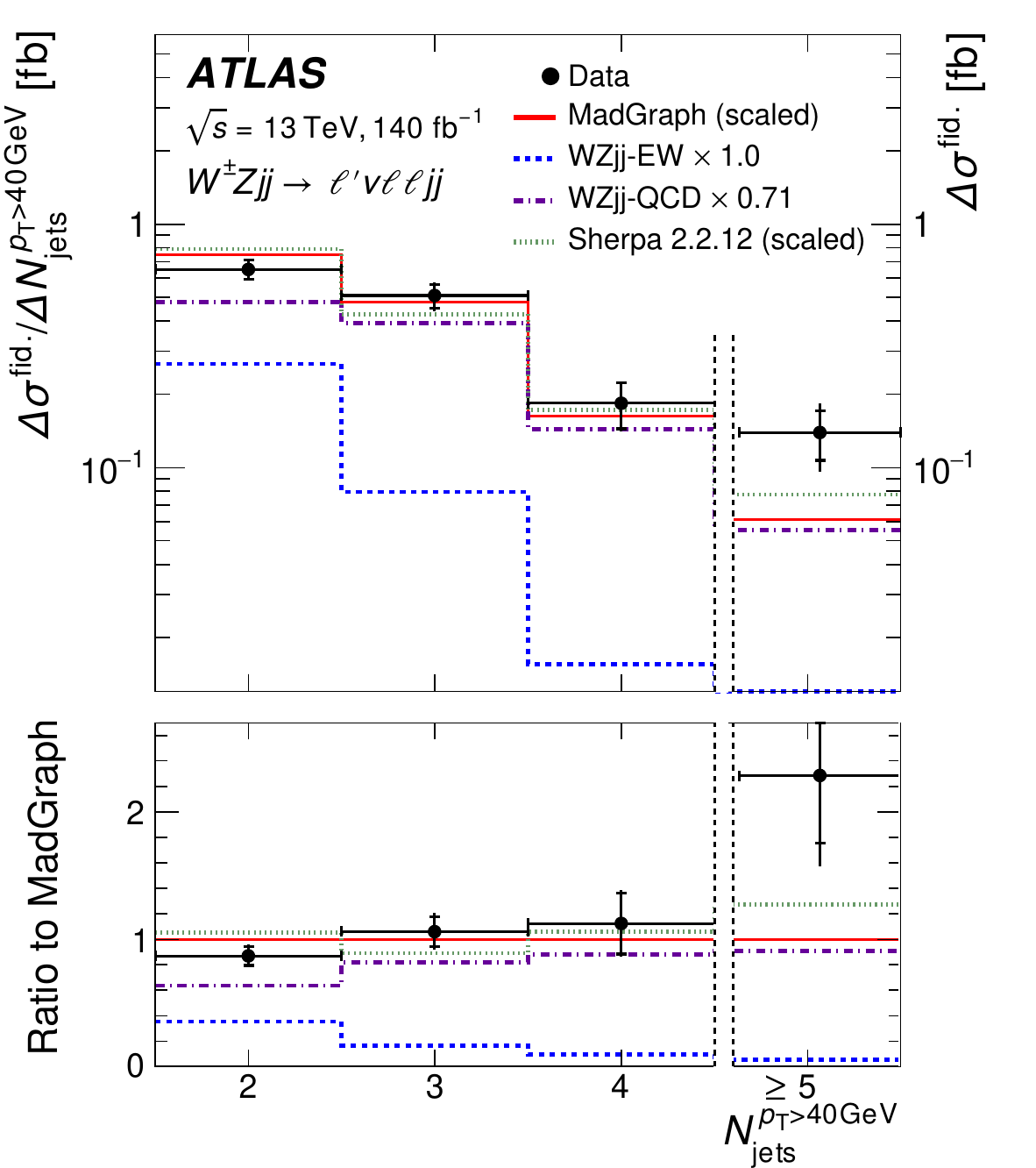}}
\subfloat[]{\includegraphics[width=0.45\textwidth]{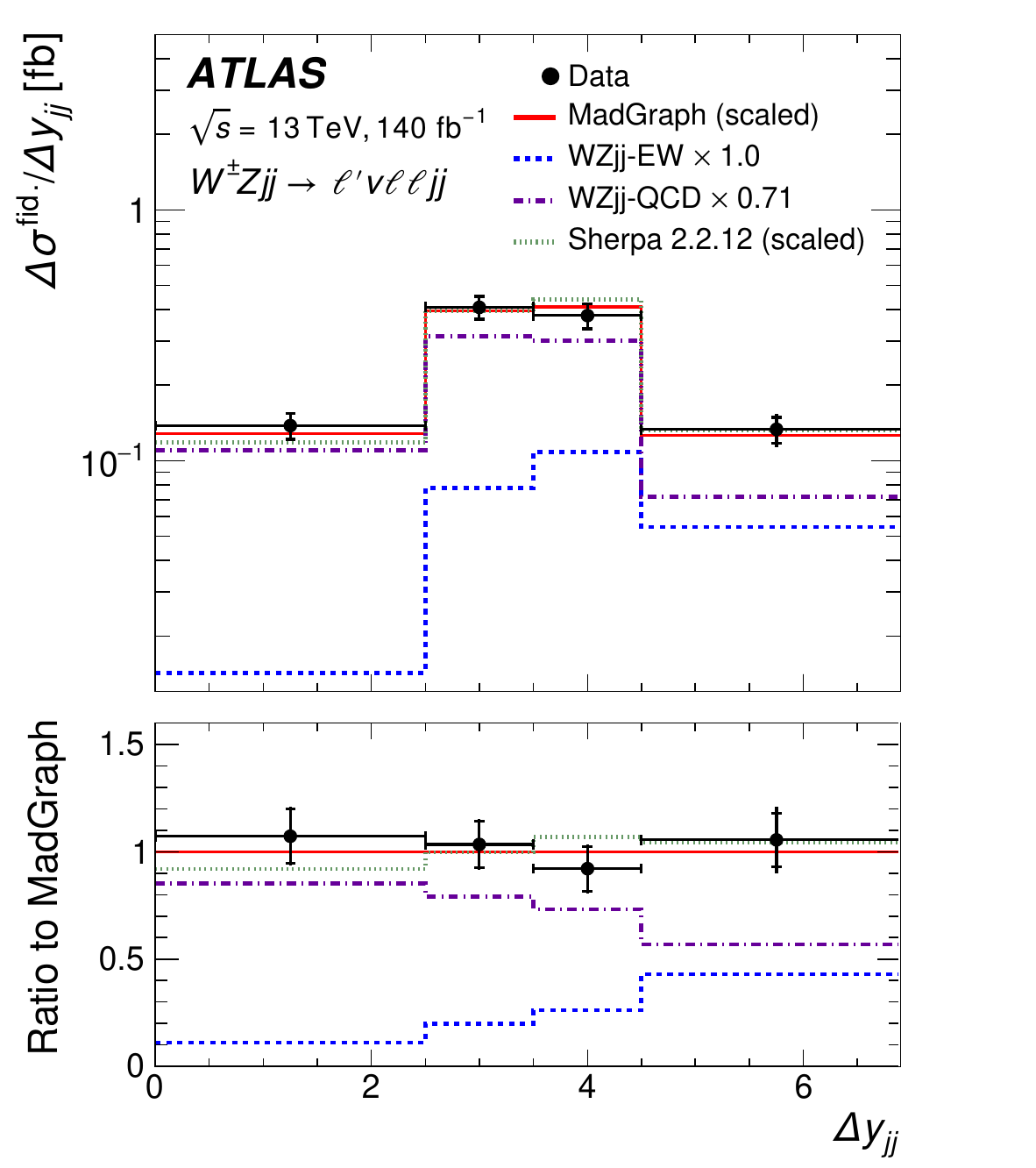}}\\
\subfloat[]{\includegraphics[width=0.45\textwidth]{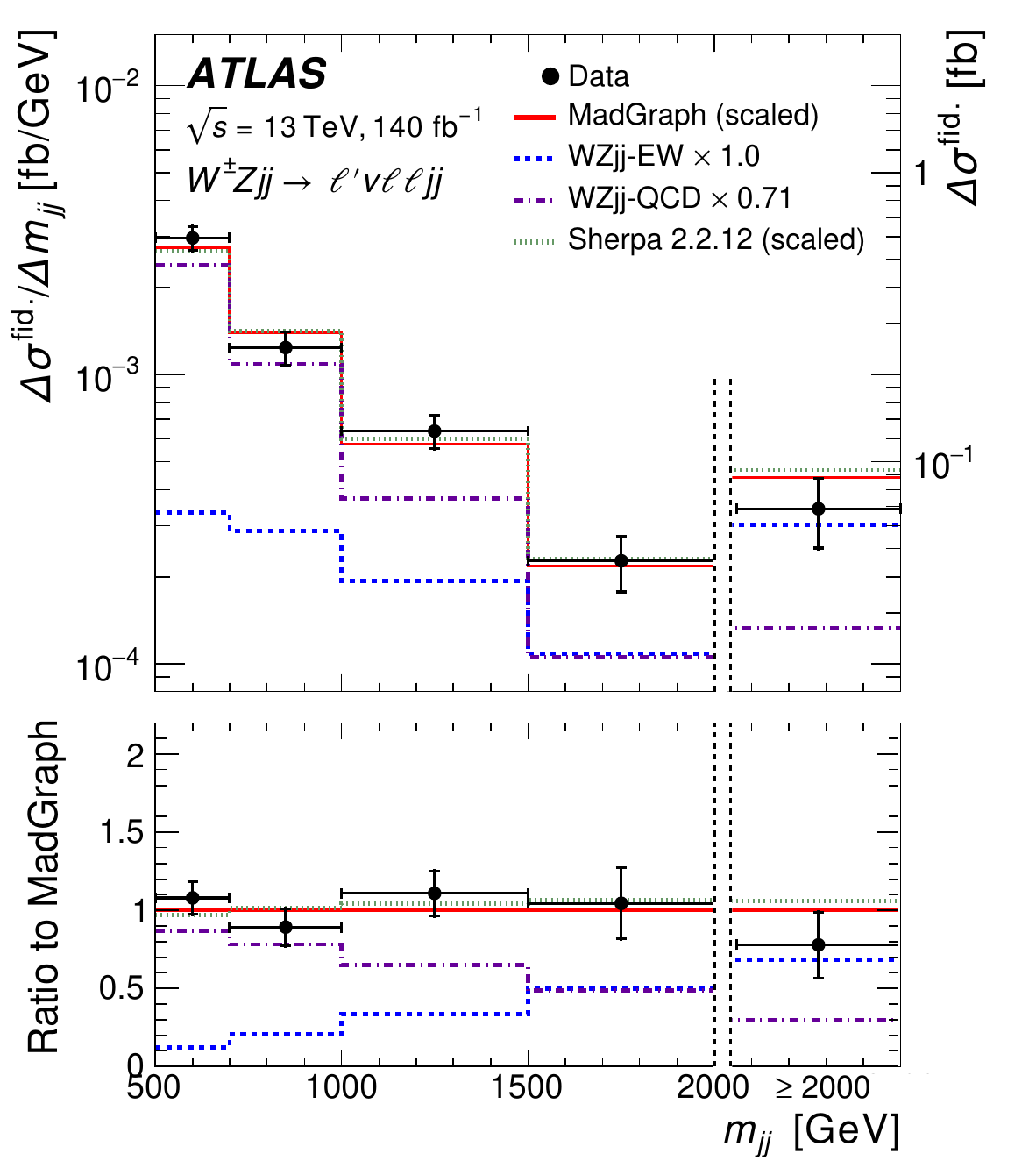}}
\subfloat[]{\includegraphics[width=0.45\textwidth]{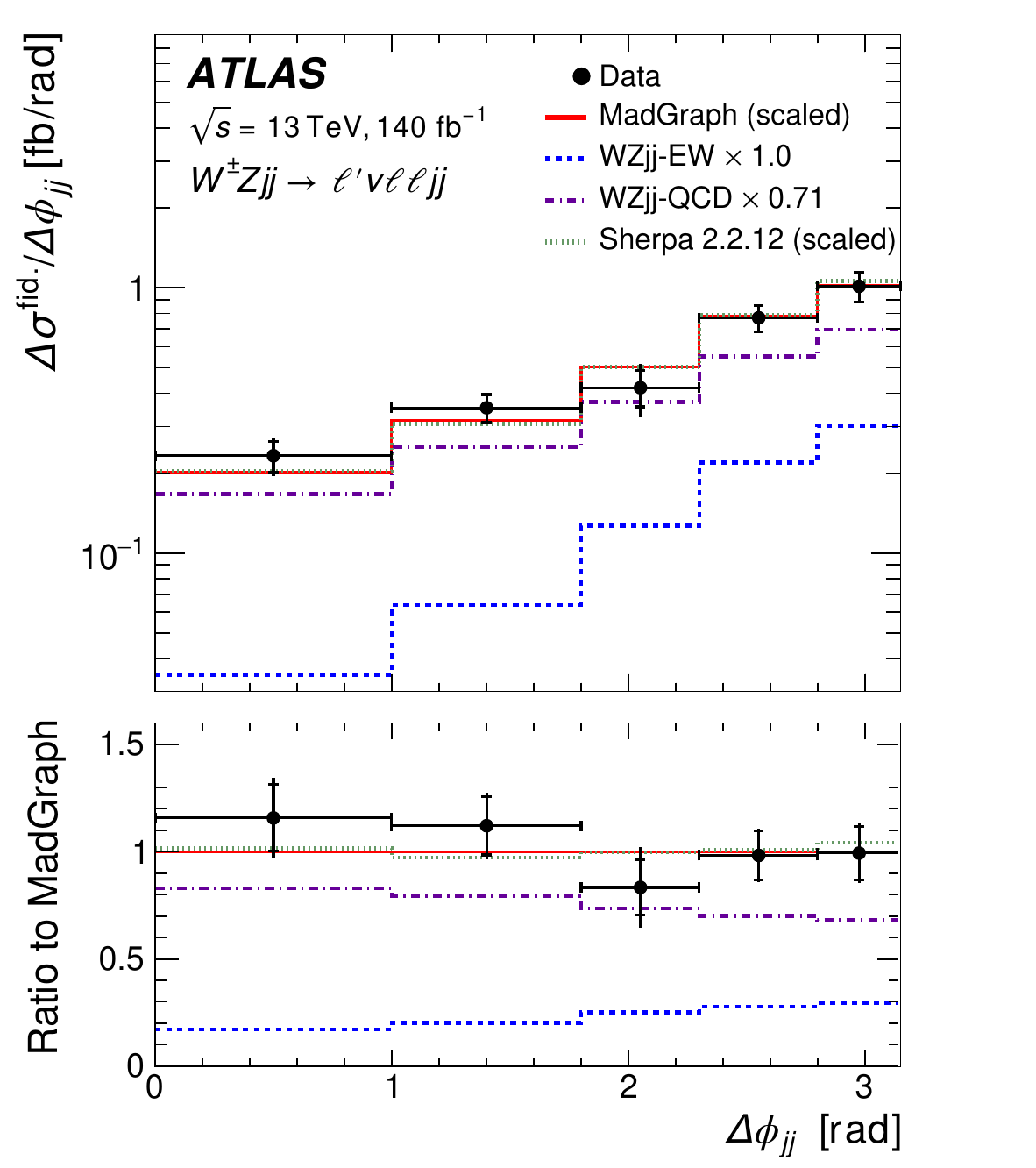}}
\caption{The measured \wzjj differential cross-section in the VBS fiducial phase space as a function of (a) the exclusive jet multiplicity of jets with $p_\textrm{T} > 40$~\GeV, \njetsf, (b) the absolute difference in rapidity between the two tagging jets \dYjj, (c) the invariant mass of the tagging jets \mjj, and (d) the azimuthal angle between the two tagging jets \dPhijj.
The inner and outer error bars on the data points represent the statistical and total uncertainties, respectively.
The measurements are compared with the sum of the rescaled \wzjjqcd and \wzjjew predictions from \MGPy8  (solid line) and \sherpa 2.2.12 (dotted line).
The \wzjjew and \wzjjqcd contributions are also represented by dashed and dashed-dotted lines, respectively.
In (a) and (c), the right $y$-axis refers to the last cross-section point, separated from the others by a vertical dashed line, as this last bin is integrated up to the maximum value reached in the phase space.
The lower panels show the ratios of the data to the predictions from \MGPy8.
}
\label{fig:diffXSection:dYjj_mjj}
\end{figure}

\begin{figure}[t]
\centering
\subfloat[]{\includegraphics[width=0.45\textwidth]{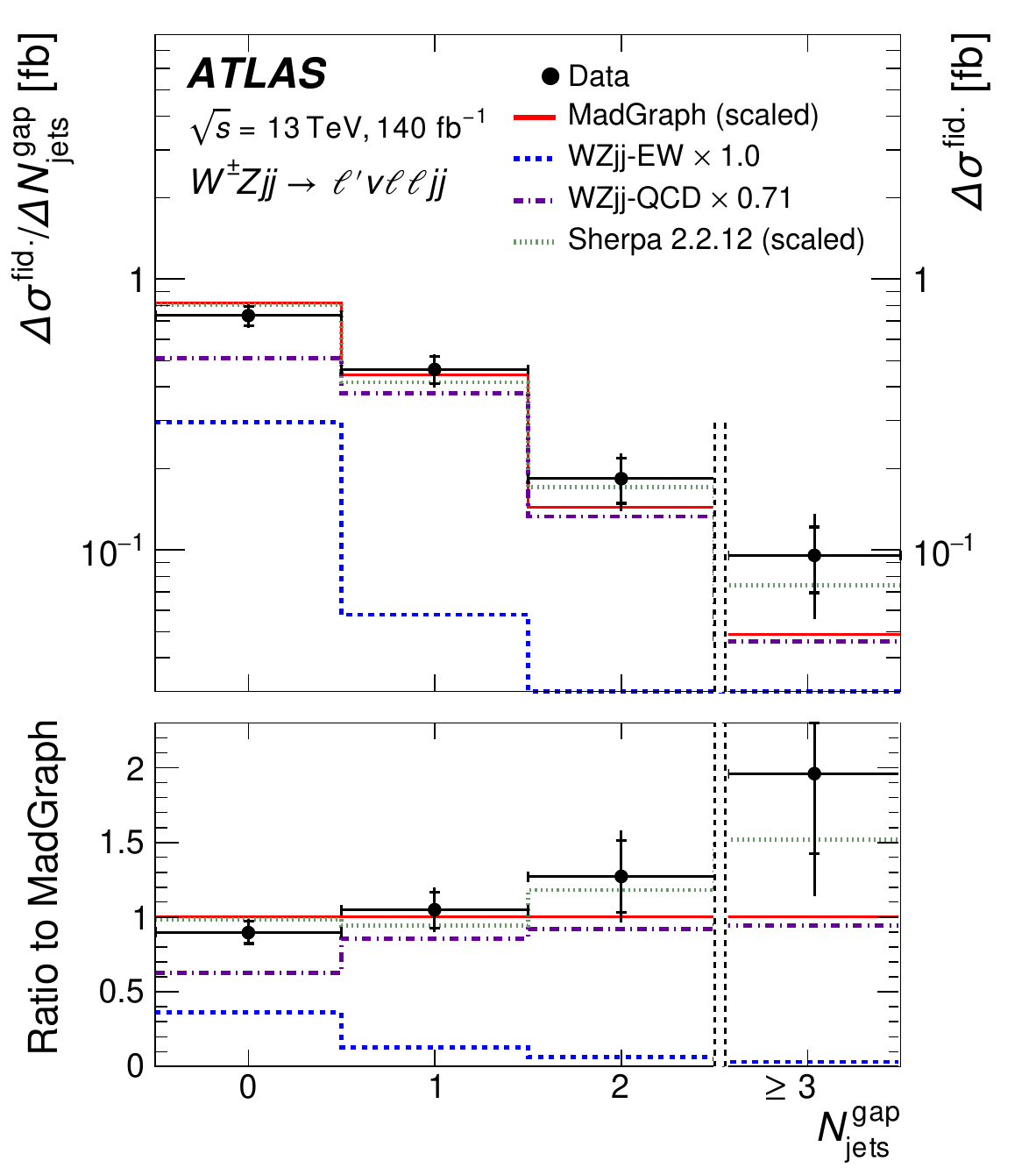}}
\subfloat[]{\includegraphics[width=0.45\textwidth]{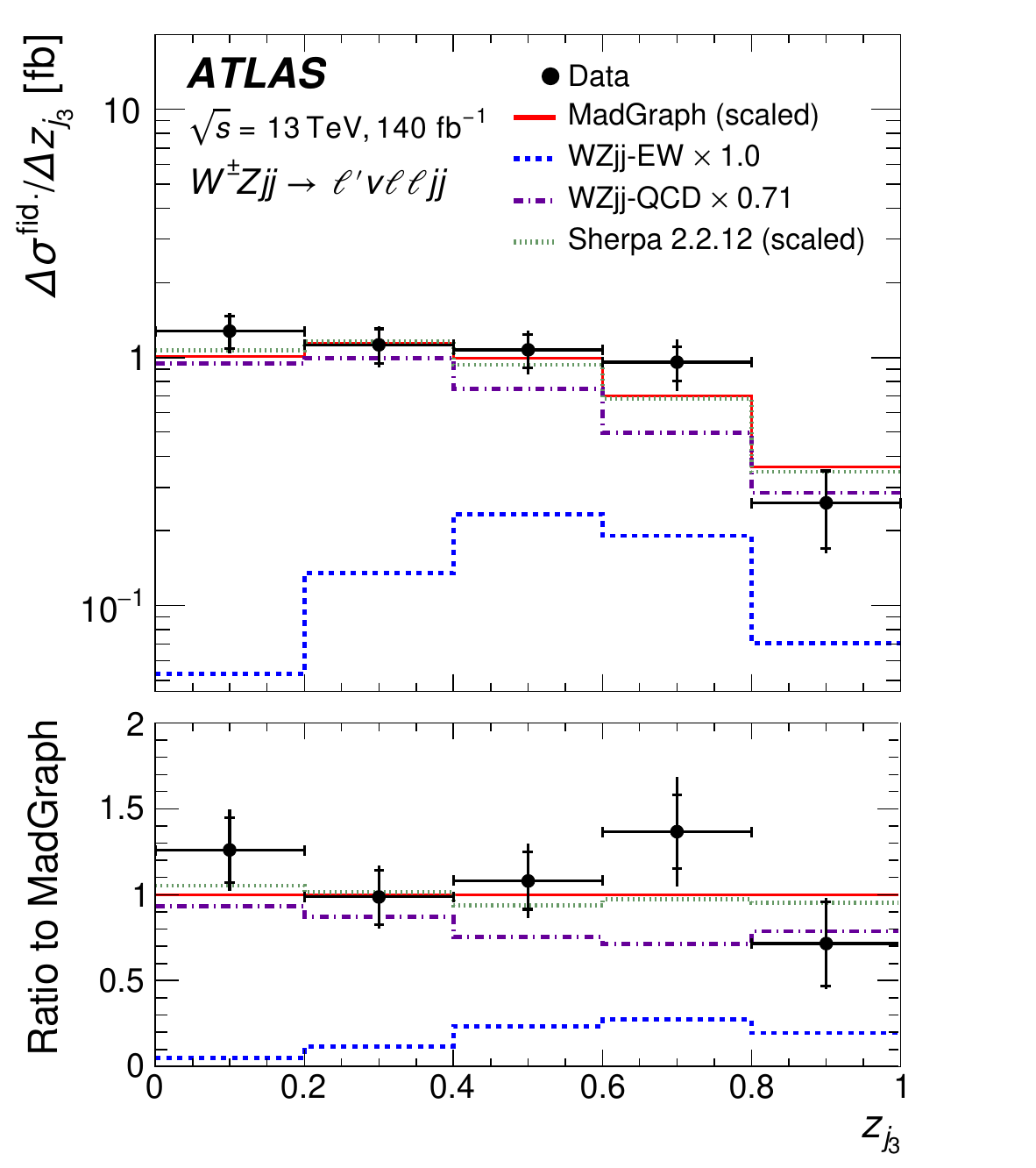}}
\caption{The measured \wzjj differential cross-section in the VBS fiducial phase space as a function of (a) \NjGap\, the exclusive jet multiplicity of jets with $p_\textrm{T} > 25$~\GeV\ in the gap between the two tagging jets, and (b) \zj3  the Zeppenfeld variable.
The inner and outer error bars on the data points represent the statistical and total uncertainties, respectively.
The measurements are compared with the sum of the rescaled \wzjjqcd and \wzjjew predictions from \MGPy8  (solid line) and \sherpa 2.2.12 (dotted line).
The \wzjjew and \wzjjqcd contributions from \MGPy8 are also represented separately by dashed and dashed-dotted lines, respectively.
In (a) the right $y$-axis refers to the last cross-section point, separated from the others by a vertical dashed line, as this last bin is integrated up to the maximum value reached in the phase space.
The lower panels show the ratios of the data to the predictions from \MGPy8.
}
\label{fig:diffXSection:3rdjet}
\end{figure}
 
\begin{figure}[t]
\centering
\includegraphics[width=0.45\textwidth]{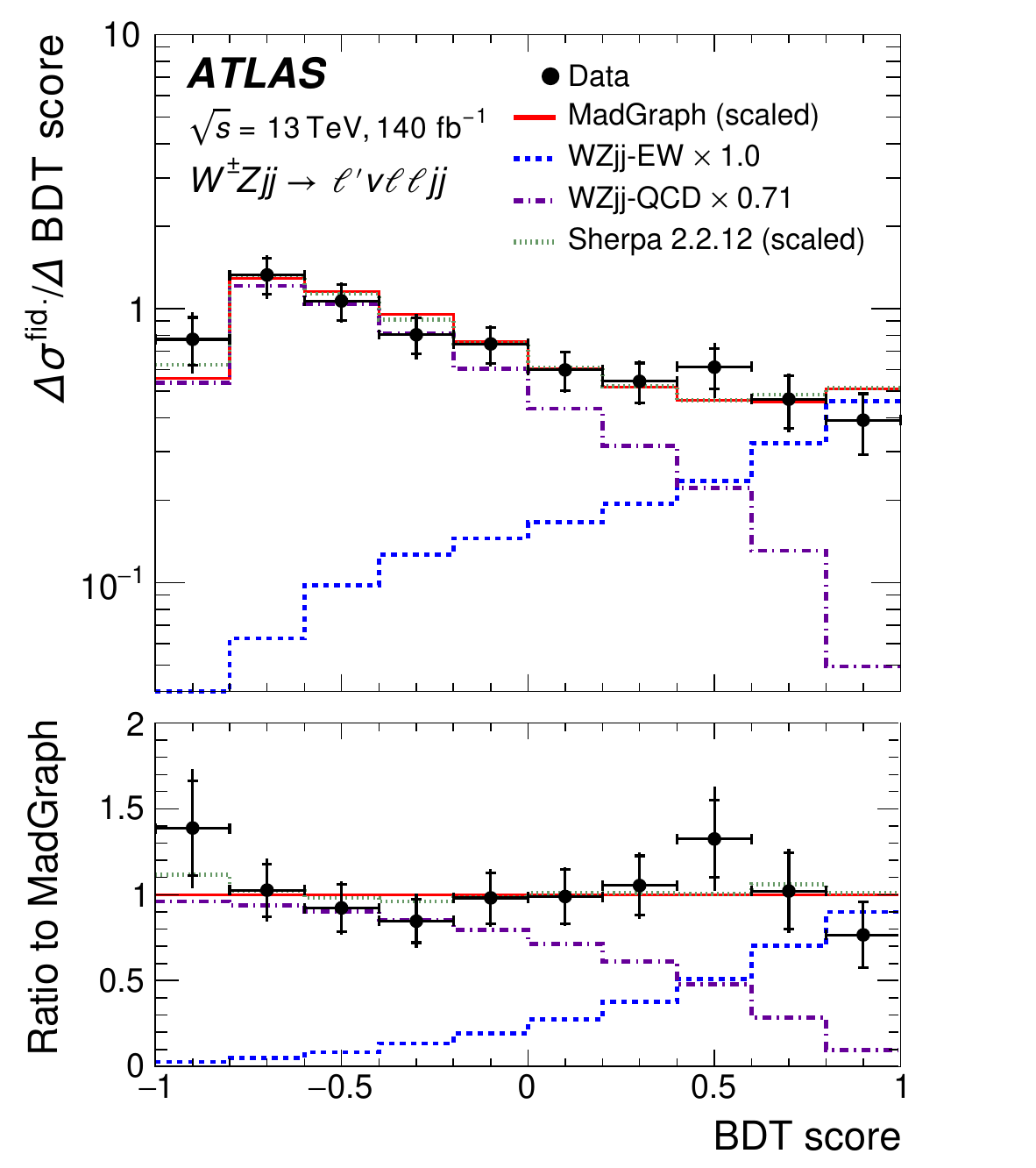}
\caption{The measured \wzjj differential cross-section in the VBS fiducial phase space as a function of the \BDTscore.
The inner and outer error bars on the data points represent the statistical and total uncertainties, respectively.
The measurements are compared with the sum of the rescaled \wzjjqcd and \wzjjew predictions from \MGPy8  (solid line) and \sherpa 2.2.12 (dotted line).
The \wzjjew and \wzjjqcd contributions from \MGPy8 are also represented separately by dashed and dashed-dotted lines, respectively.
The lower panels show the ratios of the data to the predictions from \MGPy8.
}
\label{fig:diffXSection:BDT}
\end{figure}


\clearpage
\section{Limits on anomalous quartic gauge couplings}
\label{sec:EFT_limits}


The \wzjjew production can be sensitive to effects beyond the SM affecting the quartic interactions of weak bosons.
The measured \wzjj events are therefore used to search for anomalous quartic gauge couplings (aQGC) using an EFT framework~\cite{Eboli:2006wa,Eboli:2016kko}.
In this model the SM Lagrangian $\mathcal{L}_{\mathrm{SM}}$ is extended in an effective Lagrangian $\mathcal{L}_{\mathrm{eff}}$ adding higher order operators and their respective Wilson coefficients as:

\begin{equation} \label{eq:EFT_Lagrangian}
\mathcal{L}_{\mathrm{eff}} = \mathcal{L}_{\mathrm{SM}}+\sum_i \frac{f_i^{(6)}}{\Lambda_i^2} O_i^{(6)}+\sum_j \frac{f_i^{(8)}}{\Lambda_i^4} O_j^{(8)}+\ldots \; ,
\end{equation}
 
where  $O_{i,j}^{(6),(8)}$ are dimension-6 and dimension-8 operators, respectively, and involve SM fields with respective dimensionless couplings $f_i^{(6)}$ and $f_j^{(8)}$, and $\Lambda$ is the energy scale of the new processes.
In this study only dimension-8 operators are considered and all dimension-6 couplings, affecting triple gauge boson couplings, are assumed to be equal to zero.
Dimension-6 couplings are indeed already constrained by other processes, and especially by inclusive diboson production~\cite{ATL-PHYS-PUB-2022-037,CMS:2019efc}.
The effect of dimension-6 operators in VBS processes is of interest on its own~\cite{Bellan:2021dcy,Ethier:2021ydt}, but is not studied here.
Nine independent charge-conjugate and parity conserving dimension-8 operators are considered~\cite{Eboli:2006wa}.
The  $O_{\mathrm{S}0,\, 1,\, 2}$ operators are constructed from the covariant derivative of the Higgs doublet.
The $O_{\mathrm{T}0,\, 1,\, 2}$ operators are constructed from the $\mathrm{SU}_{\mathrm{L}}(2)$ gauge fields.
The mixed operators $O_{\mathrm{M}0,\, 1,\, 7}$ involve the  $\mathrm{SU}_{\mathrm{L}}(2)$ gauge fields and the Higgs doublet.
Because the operators  $O_{\mathrm{S}0}$ and  $O_{\mathrm{S}2}$ are Hermitian conjugates of each other, they are varied simultaneously, with equal coefficient values $f_{\mathrm{S}0} = f_{\mathrm{S}2} = f_{\mathrm{S}02}$.

The squared scattering amplitude of the effective field theory prediction for \wzjj production can be written as:
 
\begin{equation}
\left| A_{\mathrm{SM}}+\sum_i c_i A_i \right|^2 = {\left| A_{\mathrm{SM}} \right| }^2 + \sum_i c_i \, 2 \operatorname{Re}(A^*_{\mathrm{SM}}A_i)+ \sum_i c_i^2 |A_i|^2 +\sum_{ij,i\neq j} c_i c_j \, 2 \operatorname{Re}(A_{i} A^*_{j}) \; ,
\label{eq:EFT}
\end{equation}
 
where $c_i = f_j^{(8)} / \Lambda^4$, $A_{\mathrm{SM}}$ is the SM scattering amplitude,  $\sum_i c_i \, 2 \operatorname{Re}(A^*_{\mathrm{SM}} A_i)$ is the amplitude of the interference term between the SM and the dimension-8 operators, $\sum_i {c_i}^2 |A_i|^2$ is the pure dimension-8 contribution, and $\sum_{ij,i\neq j} c_i c_j 2 \operatorname{Re}(A_i A^*_j)$ is the amplitude of interferences between two dimension-8 operators, called cross terms.
The different terms are simulated separately using \MGNLO[2.6.5] interfaced with \PYTHIA[8.240], providing individual MC samples.
The same PDF sets and PS modelling as for \wzjj SM events and detailed in Section~\ref{sec:samples} are used.
Generated events corresponding to a given value of the EFT coefficient $c_i$, or $c_i$ and $c_j$ for cross terms, are obtained by multiplying the respective MC samples by the coefficient value and adding them together.

As non-zero aQGC contributions will enhance the production cross-section of \wzjjew events at large diboson invariant masses, a two-dimensional combination of the \BDTscore, separating \wzjjew from \wzjjqcd events, and \mtwz observables is used to look for dimension-8 EFT contributions.
Four bins in \BDTscore ($[-1,-0.25,0.17,0.72,1]$) and five bins in \mtwz ($[0,400,750,1050,1350,\infty]$~\GeV) are used and arranged in a one-dimensional histogram of 20 statistically independent bins,  as represented in Figure~\ref{fig:2D_EFTTemplate}.
The bin boundaries are optimised to obtain the best expected limits when no unitarisation cut-off are applied.
The distribution is used to define an extended likelihood function, adding the same control regions as defined in Section~\ref{sec:meth_inclusive}.
Experimental and theory uncertainties, as discussed in section~\ref{sec:systematics}, are included as Gaussian-constrained nuisance parameters.
A profile-likelihood ratio test statistic is constructed to estimate the confidence intervals for a given $c_i$.
For each individual $c_i$, or pair $(c_i, c_j)$ for two-dimensional limits, a maximum-likelihood fit to data is performed by setting other coefficients to zero.
The same fit parameterisation as described in Section~\ref{sec:meth_inclusive} is used.

\begin{figure}[t]
\centering
\includegraphics[scale=0.6]{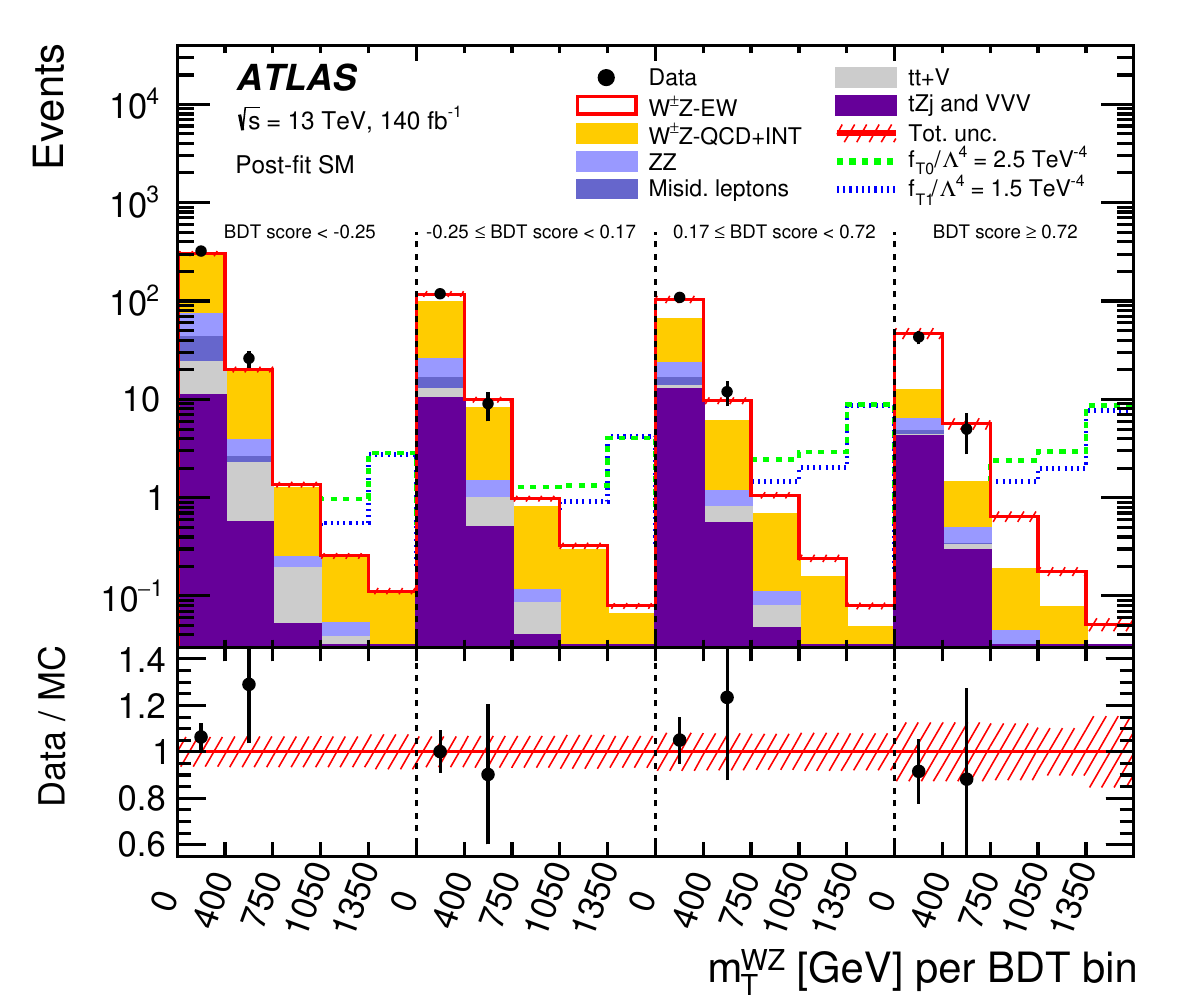}
\caption{Detector-level one-dimension distribution of the two-dimensional combination of \BDTscore and \mtwz observables used to obtain limits on EFT coefficients.
The SM predicted event yields resulting from the SM-only fit to data are shown.
The uncertainty band around the MC expectation includes all systematic uncertainties as obtained from the fit.
The contribution of the $O_{\mathrm{T}0}$ and $O_{\mathrm{T}1}$ operators with illustrative values of the corresponding Wilson coefficients are also represented.}
\label{fig:2D_EFTTemplate}
\end{figure}

\begin{table}[t]
 
\caption{
Expected and observed 95\% CL intervals on the Wilson coefficients of the different dimension-8 operators, with no unitarisation procedure.
The notation $\mathrm{S}02$ is used to indicate that the coefficients corresponding to operators  $O_{\mathrm{S}0}$ and $O_{\mathrm{S}2}$ are assigned the same value.
}
\label{tab:limitsNoCl}
\centering
\begin{tabular}{c c c c c c c c c}
\toprule
& Expected  &  Observed  &  \\
& [\TeV$^{-4}$]& [\TeV$^{-4}$] &\\\hline
$f_{\mathrm{T}0}/{\Lambda}^4$ & [-0.80, 0.80] & [-0.57, 0.56]  &   \\
$f_{\mathrm{T}1}/{\Lambda}^4$ & [-0.52, 0.49] & [-0.39, 0.35] & \\
$f_{\mathrm{T}2}/{\Lambda}^4$ & [-1.6, 1.4] & [-1.2, 1.0] &  \\
$f_{\mathrm{M}0}/{\Lambda}^4$ & [-8.3, 8.3] & [-5.8, 5.6] &   \\
$f_{\mathrm{M}1}/{\Lambda}^4$ & [-12.3, 12.2] & [-8.6, 8.5] &   \\
$f_{\mathrm{M}7}/{\Lambda}^4$ & [-16.2, 16.2] & [-11.3, 11.3]  &  \\
$f_{\mathrm{S}02}/{\Lambda}^4$ & [-14.2, 14.2] & [-10.4, 10.4]  &   \\
$f_{\mathrm{S}1}/{\Lambda}^4$ & [-42, 41] & [-30, 30]  &   \\
\bottomrule
\end{tabular}

\end{table}

No deviation with respect to the SM predictions is observed and the expected and observed $95$\%  confidence level (CL) lower and upper limits on the given Wilson coefficients are presented in Table~\ref{tab:limitsNoCl}.
Coefficients associated to the $O_{\mathrm{T}0}$ and $O_{\mathrm{T}1}$ operators are the most tightly constraint.
The limits are dominated by the pure dimension-8 contribution in Eq.~(\ref{eq:EFT}).
These limits are similar to those obtained by the CMS experiment also using leptonic decay modes of \wzjj events~\cite{CMS:2020gfh}.
Two-dimensional limits at $95$\% CL on the pair of $O_{\mathrm{T}0}$ and $O_{\mathrm{T}1}$ operators are also obtained and presented in Figure~\ref{fig:2D_exp_plots_reco}.
No unitarisation procedure is applied to obtain these results.

\begin{figure}[t]
\centering
\includegraphics[scale=0.4]{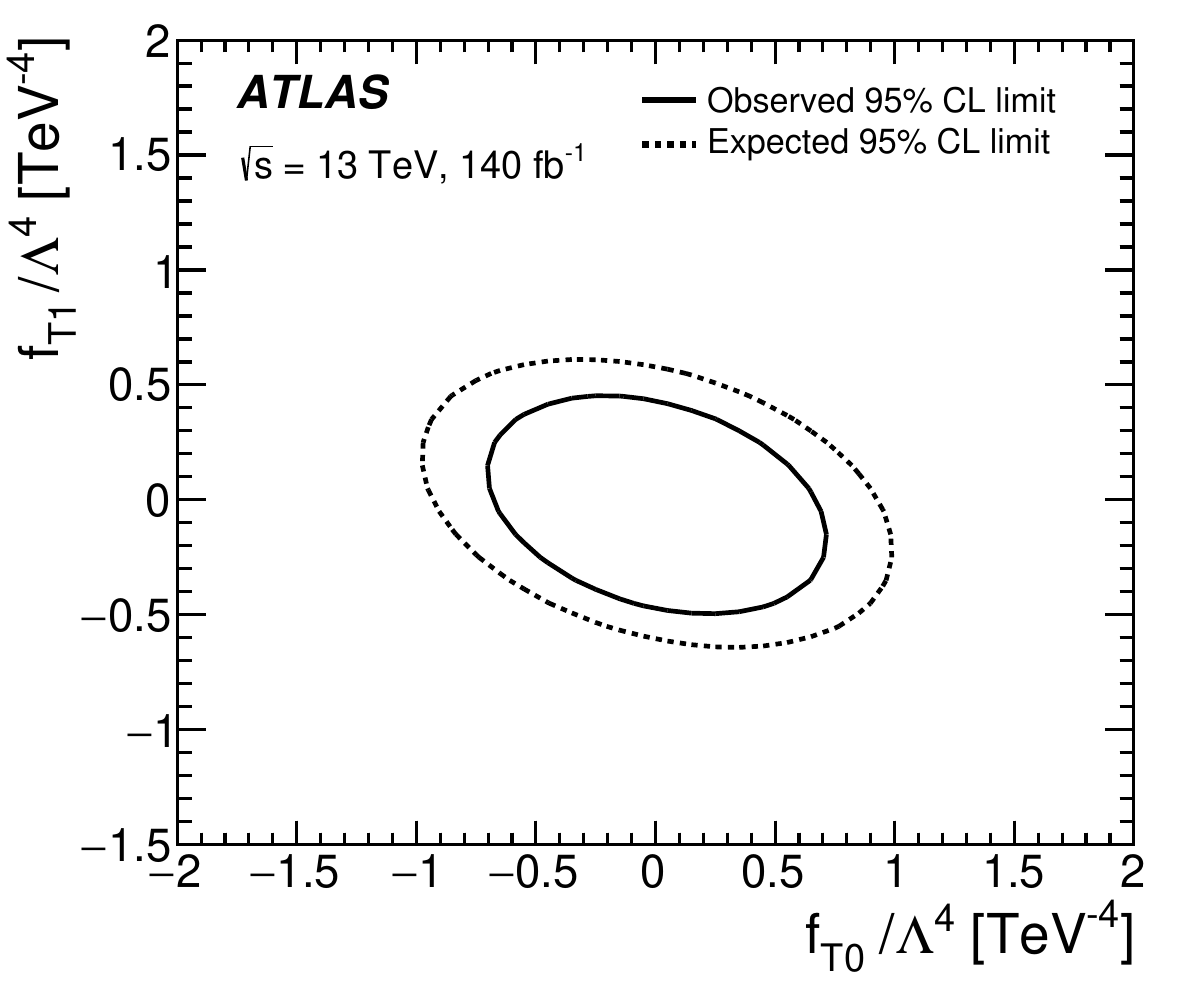}
\caption{Two-dimensional expected (dashed line) and observed (solid line) 95\% CL intervals on Wilson coefficients corresponding to the pair of  $O_{\mathrm{T}0}$ and $O_{\mathrm{T}1}$ operators.
No unitarisation procedure is used.
}
\label{fig:2D_exp_plots_reco}
\end{figure}

The EFT is not a complete model and the presence of non-zero dimension-8 operators violates tree-level unitarity at sufficiently high energy.
More physical limits are obtained by removing the EFT contribution above the unitarity limit and keeping the SM prediction for all $WZ$ invariant masses, even above the unitarity limit.
The unitarity limits from Ref.~\cite{Almeida:2020ylr} are used with only one non-zero Wilson coefficient.
These unitarity limits are presented in Figure~\ref{fig:clipping_T} for the Wilson coefficients  $f_{\mathrm{T}0}$ and $f_{\mathrm{T}1}$, together with the evolution of the expected and observed $95$\% CL intervals as a function of the cut-off scale $m_{WZ}$ used in the unitarisation procedure.

Table~\ref{tab:limitsWithCl} shows the individual $95$\% CL intervals of each Wilson coefficients obtained when applying a unitarisation cut-off at the unitarity bound.
For the $O_{\mathrm{S}1}$ operator no crossing with the unitarity bound was found in the scanned region above $600$~\GeV and therefore no limits are reported.

\begin{figure}[t]
\centering
\subfloat[]{\includegraphics[scale=0.4]{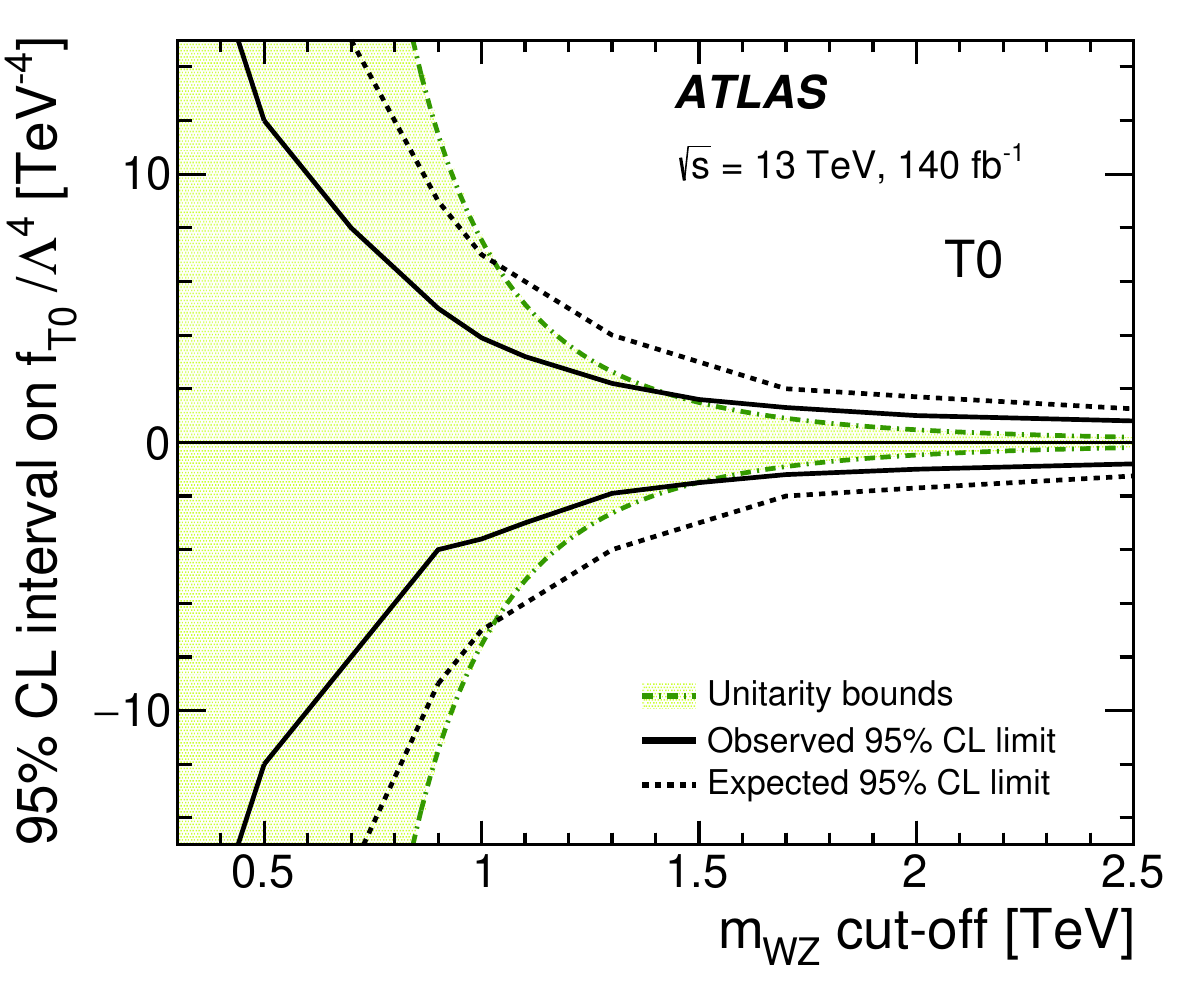}}
\subfloat[]{\includegraphics[scale=0.4]{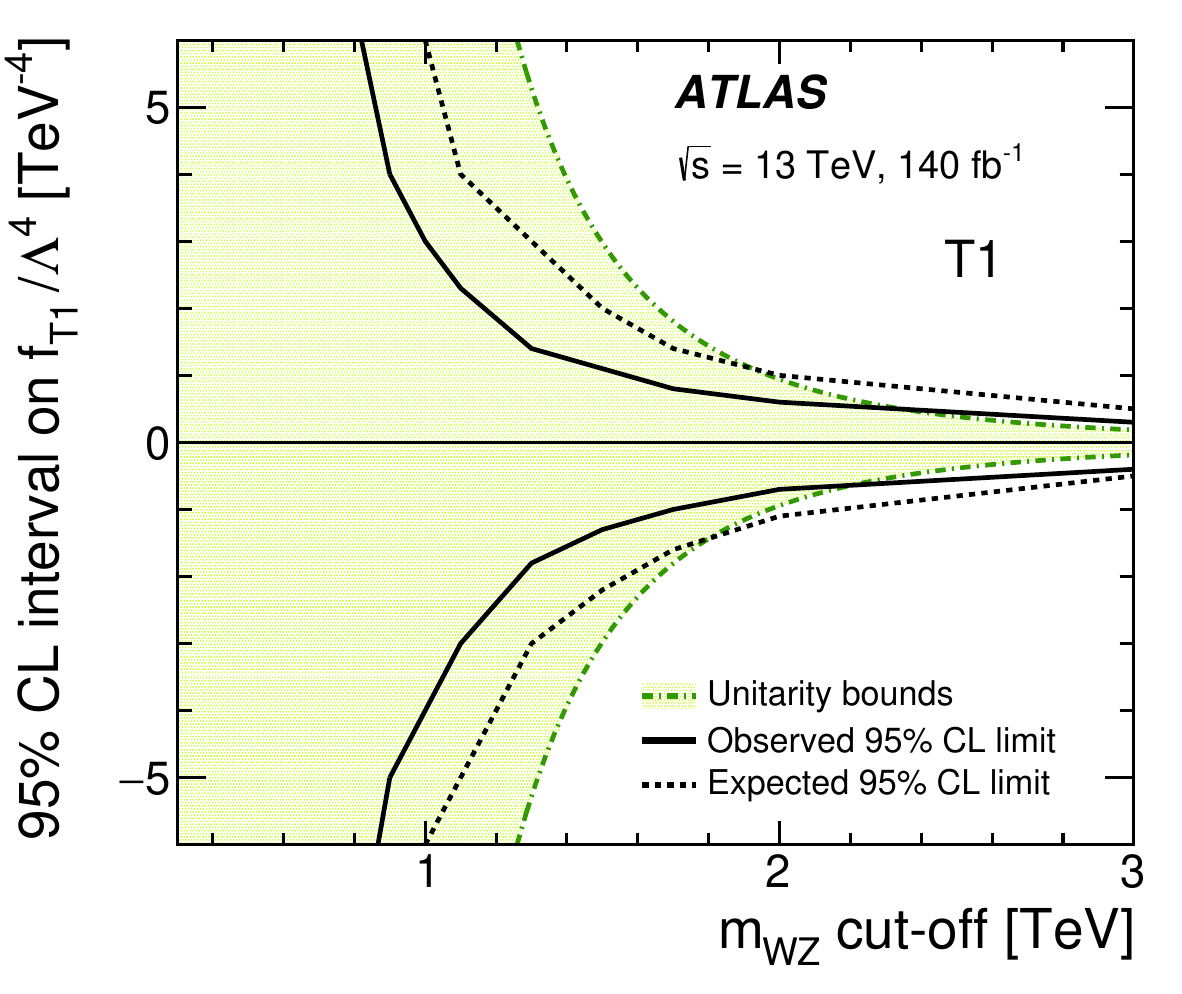}}
\caption{Evolution  as a function of the cut-off scale of the individual expected (dashed lines) and observed (solid lines) limits at 95\% CL on the Wilson coefficients corresponding to the operators (a) $O_{\mathrm{T}0}$ and (b) $O_{\mathrm{T}1}$. The shaded area represents the unitarity allowed region.
The unitarity bounds (dotted-dashed line) for each operator are defined for one non-zero Wilson coefficient following Ref.~\cite{Almeida:2020ylr}.
}
\label{fig:clipping_T}
\end{figure}

\begin{table}[t]
\caption{
Expected and observed 95\% CL intervals on the Wilson coefficients of the different dimension-8 operators, with a unitarisation cut-off set at the unitary bound, ie. where the unitary bound and the experimental bound cross.
The notation $\mathrm{S}02$ is used to indicate that the coefficients corresponding to operators  $O_{\mathrm{S}0}$ and $O_{\mathrm{S}2}$ are assigned the same value.
}
\label{tab:limitsWithCl}
\centering
 
\begin{tabular}{c c c c c c c c c}
\toprule
& Expected  &  Observed &  \\
& [\TeV$^{-4}$] & [\TeV$^{-4}$] &\\\hline
$f_{\mathrm{T}0}/{\Lambda}^4$ & [-7.0, 7.0] & [-1.5, 1.6]  &   \\
$f_{\mathrm{T}1}/{\Lambda}^4$ & [-1.1, 1.0] & [-0.7, 0.6] & \\
$f_{\mathrm{T}2}/{\Lambda}^4$ & [-12, 6] & [-2.4, 1.8] &  \\
$f_{\mathrm{M}0}/{\Lambda}^4$ & [-60, 60] & [-12, 12] &   \\
$f_{\mathrm{M}1}/{\Lambda}^4$ & [-32, 32] & [-15, 15] &   \\
$f_{\mathrm{M}7}/{\Lambda}^4$ & [-30, 30] & [-15, 15]  &  \\
$f_{\mathrm{S}02}/{\Lambda}^4$ & [-41, 41] & [-18, 18]  &   \\
$f_{\mathrm{S}1}/{\Lambda}^4$ & --- & ---  &   \\
\bottomrule
\end{tabular}
 
\end{table}


\FloatBarrier

\section{Conclusion}
\label{sec:conclusion}

Measurements of integrated and differential cross-sections of electroweak production of a \wz\ pair in association with two jets and measurements of its  production cross-section in $\sqrt{s} = 13$~\TeV\ $pp$ collisions at the LHC are presented.
The data were collected with the ATLAS detector and correspond to an integrated luminosity of $140~\ifb$.
The measurements use leptonic decays of the gauge bosons into electrons or muons and are performed in a fiducial phase space approximating the detector acceptance that increases the sensitivity to \wzjj\ electroweak production modes.
 
The measured integrated fiducial cross-sections for a single leptonic decay mode of electroweak and inclusive \wzjj production are  $\sigma_{WZjj\mathrm{-EW}  \rightarrow \ell^{'} \nu \ell \ell jj} = 0.368 \; \pm  0.037 \,(\mathrm{stat.}) \; \pm 0.059  \,(\mathrm{syst.}) \; \pm 0.003 \,(\mathrm{lumi.})  \; \mathrm{fb}$ and  $\sigma_{WZjj  \rightarrow \ell^{'} \nu \ell \ell jj} = 1.462 \; \pm 0.063 \,(\mathrm{stat.}) \; \pm 0.118 \,(\mathrm{syst.})  \; \pm 0.012 \,(\mathrm{lumi.}) \; \mathrm{fb}$, respectively.
This is the most precise measurement to date of the \wzjjew production cross-section.
The electroweak \wzjj production cross-section is found to agree with the \LO SM prediction of  $0.37 \pm 0.03$~fb as calculated with the \MGPy8 MC event generator.
However, the inclusive \wzjj production cross-section is found to be smaller than the SM prediction from the \MGPy8 generator of $1.91 \pm 0.17$~fb.
The reason for this is that the \wzjjqcd integrated cross-section is measured to be lower than the prediction from \MGPy8 by a factor of $0.7$, with a compatibility of $1.8\sigma$ between the measurement and the prediction.
 
Differential cross-sections of electroweak and strong \wzjj production are measured for the first time separately for events with exactly two jets or with more than two jets, as well as in three bins of the invariant mass of the two tagging jets.
For electroweak \wzjj production, MC predictions from both \MGPy8 or \SHERPA[2.2.12] are found to agree with the measured cross-sections within two standard deviations in all sub-categories of signal region events.
The strong  \wzjj production cross-section is however found to be largely mis-modelled by MC predictions from both \MGPy8 or \sherpa~2.2.12 for events with exactly two jets of $\pt > 25~\GeV$ or with $500 < \mjj < 1300~\GeV$.
Differential cross-sections of \wzjj production, including both the strong and electroweak processes, are also measured in the same fiducial phase space as a function of several kinematic observables.
 
Finally, the measured \wzjj events are also used to constrain anomalous quartic gauge couplings by extracting $95$\% confidence level intervals on the dimension-$8$ operators T0, T1, T2, M0, M1, M7, S02 and S1, with and without consideration of tree-level unitarity violation.
These constraints are similar to those obtained by the CMS Collaboration using the \wzjj final state.


\section*{Acknowledgements}


We thank CERN for the very successful operation of the LHC and its injectors, as well as the support staff at
CERN and at our institutions worldwide without whom ATLAS could not be operated efficiently.
 
The crucial computing support from all WLCG partners is acknowledged gratefully, in particular from CERN, the ATLAS Tier-1 facilities at TRIUMF/SFU (Canada), NDGF (Denmark, Norway, Sweden), CC-IN2P3 (France), KIT/GridKA (Germany), INFN-CNAF (Italy), NL-T1 (Netherlands), PIC (Spain), RAL (UK) and BNL (USA), the Tier-2 facilities worldwide and large non-WLCG resource providers. Major contributors of computing resources are listed in Ref.~\cite{ATL-SOFT-PUB-2023-001}.
 
We gratefully acknowledge the support of ANPCyT, Argentina; YerPhI, Armenia; ARC, Australia; BMWFW and FWF, Austria; ANAS, Azerbaijan; CNPq and FAPESP, Brazil; NSERC, NRC and CFI, Canada; CERN; ANID, Chile; CAS, MOST and NSFC, China; Minciencias, Colombia; MEYS CR, Czech Republic; DNRF and DNSRC, Denmark; IN2P3-CNRS and CEA-DRF/IRFU, France; SRNSFG, Georgia; BMBF, HGF and MPG, Germany; GSRI, Greece; RGC and Hong Kong SAR, China; ISF and Benoziyo Center, Israel; INFN, Italy; MEXT and JSPS, Japan; CNRST, Morocco; NWO, Netherlands; RCN, Norway; MNiSW, Poland; FCT, Portugal; MNE/IFA, Romania; MESTD, Serbia; MSSR, Slovakia; ARRS and MIZ\v{S}, Slovenia; DSI/NRF, South Africa; MICINN, Spain; SRC and Wallenberg Foundation, Sweden; SERI, SNSF and Cantons of Bern and Geneva, Switzerland; MOST, Taipei; TENMAK, T\"urkiye; STFC, United Kingdom; DOE and NSF, United States of America.
 
Individual groups and members have received support from BCKDF, CANARIE, CRC and DRAC, Canada; PRIMUS 21/SCI/017, CERN-CZ and FORTE, Czech Republic; COST, ERC, ERDF, Horizon 2020, ICSC-NextGenerationEU and Marie Sk{\l}odowska-Curie Actions, European Union; Investissements d'Avenir Labex, Investissements d'Avenir Idex and ANR, France; DFG and AvH Foundation, Germany; Herakleitos, Thales and Aristeia programmes co-financed by EU-ESF and the Greek NSRF, Greece; BSF-NSF and MINERVA, Israel; Norwegian Financial Mechanism 2014-2021, Norway; NCN and NAWA, Poland; La Caixa Banking Foundation, CERCA Programme Generalitat de Catalunya and PROMETEO and GenT Programmes Generalitat Valenciana, Spain; G\"{o}ran Gustafssons Stiftelse, Sweden; The Royal Society and Leverhulme Trust, United Kingdom.
 
In addition, individual members wish to acknowledge support from CERN: European Organization for Nuclear Research (CERN PJAS); Chile: Agencia Nacional de Investigaci\'on y Desarrollo (FONDECYT 1190886, FONDECYT 1210400, FONDECYT 1230812, FONDECYT 1230987); China: National Natural Science Foundation of China (NSFC - 12175119, NSFC 12275265, NSFC-12075060); Czech Republic: Czech Science Foundation (GACR - 24-11373S), Ministry of Education Youth and Sports (FORTE CZ.02.01.01/00/22\_008/0004632), PRIMUS Research Programme (PRIMUS/21/SCI/017); European Union: European Research Council (ERC - 948254, ERC 101089007), Horizon 2020 Framework Programme (MUCCA - CHIST-ERA-19-XAI-00), European Union, Future Artificial Intelligence Research (FAIR-NextGenerationEU PE00000013), Italian Center for High Performance Computing, Big Data and Quantum Computing (ICSC, NextGenerationEU); France: Agence Nationale de la Recherche (ANR-20-CE31-0013, ANR-21-CE31-0013, ANR-21-CE31-0022), Investissements d'Avenir Labex (ANR-11-LABX-0012); Germany: Baden-Württemberg Stiftung (BW Stiftung-Postdoc Eliteprogramme), Deutsche Forschungsgemeinschaft (DFG - 469666862, DFG - CR 312/5-2); Italy: Istituto Nazionale di Fisica Nucleare (ICSC, NextGenerationEU); Japan: Japan Society for the Promotion of Science (JSPS KAKENHI JP21H05085, JSPS KAKENHI JP22H01227, JSPS KAKENHI JP22H04944, JSPS KAKENHI JP22KK0227); Netherlands: Netherlands Organisation for Scientific Research (NWO Veni 2020 - VI.Veni.202.179); Norway: Research Council of Norway (RCN-314472); Poland: Polish National Agency for Academic Exchange (PPN/PPO/2020/1/00002/U/00001), Polish National Science Centre (NCN 2021/42/E/ST2/00350, NCN OPUS nr 2022/47/B/ST2/03059, NCN UMO-2019/34/E/ST2/00393, UMO-2020/37/B/ST2/01043, UMO-2021/40/C/ST2/00187, UMO-2022/47/O/ST2/00148, UMO-2023/49/B/ST2/04085); Slovenia: Slovenian Research Agency (ARIS grant J1-3010); Spain: Generalitat Valenciana (Artemisa, FEDER, IDIFEDER/2018/048), Ministry of Science and Innovation (MCIN \& NextGenEU PCI2022-135018-2, MICIN \& FEDER PID2021-125273NB, RYC2019-028510-I, RYC2020-030254-I, RYC2021-031273-I, RYC2022-038164-I), PROMETEO and GenT Programmes Generalitat Valenciana (CIDEGENT/2019/023, CIDEGENT/2019/027); Sweden: Swedish Research Council (Swedish Research Council 2023-04654, VR 2018-00482, VR 2022-03845, VR 2022-04683, VR 2023-03403, VR grant 2021-03651), Knut and Alice Wallenberg Foundation (KAW 2018.0157, KAW 2018.0458, KAW 2019.0447, KAW 2022.0358); Switzerland: Swiss National Science Foundation (SNSF - PCEFP2\_194658); United Kingdom: Leverhulme Trust (Leverhulme Trust RPG-2020-004), Royal Society (NIF-R1-231091); United States of America: U.S. Department of Energy (ECA DE-AC02-76SF00515), Neubauer Family Foundation.


\printbibliography
 
\clearpage
 
\begin{flushleft}
\hypersetup{urlcolor=black}
{\Large The ATLAS Collaboration}

\bigskip

\AtlasOrcid[0000-0002-6665-4934]{G.~Aad}$^\textrm{\scriptsize 103}$,
\AtlasOrcid[0000-0001-7616-1554]{E.~Aakvaag}$^\textrm{\scriptsize 16}$,
\AtlasOrcid[0000-0002-5888-2734]{B.~Abbott}$^\textrm{\scriptsize 121}$,
\AtlasOrcid[0000-0002-0287-5869]{S.~Abdelhameed}$^\textrm{\scriptsize 117a}$,
\AtlasOrcid[0000-0002-1002-1652]{K.~Abeling}$^\textrm{\scriptsize 55}$,
\AtlasOrcid[0000-0001-5763-2760]{N.J.~Abicht}$^\textrm{\scriptsize 49}$,
\AtlasOrcid[0000-0002-8496-9294]{S.H.~Abidi}$^\textrm{\scriptsize 29}$,
\AtlasOrcid[0009-0003-6578-220X]{M.~Aboelela}$^\textrm{\scriptsize 44}$,
\AtlasOrcid[0000-0002-9987-2292]{A.~Aboulhorma}$^\textrm{\scriptsize 35e}$,
\AtlasOrcid[0000-0001-5329-6640]{H.~Abramowicz}$^\textrm{\scriptsize 152}$
% ...
\bigskip
\\

% ...
$^{16}$Department for Physics and Technology, University of Bergen, Bergen; Norway.\\
% ...
$^{29}$Physics Department, Brookhaven National Laboratory, Upton NY; United States of America.\\
% ...
$^{35}$$^{(a)}$Facult\'e des Sciences Ain Chock, R\'eseau Universitaire de Physique des Hautes Energies - Universit\'e Hassan II, Casablanca;$^{(b)}$Facult\'{e} des Sciences, Universit\'{e} Ibn-Tofail, K\'{e}nitra;$^{(c)}$Facult\'e des Sciences Semlalia, Universit\'e Cadi Ayyad, LPHEA-Marrakech;$^{(d)}$LPMR, Facult\'e des Sciences, Universit\'e Mohamed Premier, Oujda;$^{(e)}$Facult\'e des sciences, Universit\'e Mohammed V, Rabat;$^{(f)}$Institute of Applied Physics, Mohammed VI Polytechnic University, Ben Guerir; Morocco.\\
% ...
$^{44}$Physics Department, Southern Methodist University, Dallas TX; United States of America.\\
% ...
$^{49}$Fakult\"{a}t Physik , Technische Universit{\"a}t Dortmund, Dortmund; Germany.\\
% ...
$^{55}$II. Physikalisches Institut, Georg-August-Universit\"{a}t G\"ottingen, G\"ottingen; Germany.\\
% ...
$^{103}$CPPM, Aix-Marseille Universit\'e, CNRS/IN2P3, Marseille; France.\\
% ...
$^{117}$$^{(a)}$New York University Abu Dhabi, Abu Dhabi;$^{(b)}$United Arab Emirates University, Al Ain; United Arab Emirates.\\
% ...
$^{121}$Homer L. Dodge Department of Physics and Astronomy, University of Oklahoma, Norman OK; United States of America.\\
% ...
$^{152}$Raymond and Beverly Sackler School of Physics and Astronomy, Tel Aviv University, Tel Aviv; Israel.\\
% ...
$^{a}$ Also Affiliated with an institute covered by a cooperation agreement with CERN.\\
$^{b}$ Also at An-Najah National University, Nablus; Palestine.\\
$^{c}$ Also at Borough of Manhattan Community College, City University of New York, New York NY; United States of America.\\
$^{d}$ Also at Center for Interdisciplinary Research and Innovation (CIRI-AUTH), Thessaloniki; Greece.\\
$^{e}$ Also at Centro Studi e Ricerche Enrico Fermi; Italy.\\
$^{f}$ Also at CERN, Geneva; Switzerland.\\
% ...

\end{flushleft}


\end{document}